\documentclass[twocolumn]{aastex62}

\usepackage{graphicx}
\usepackage{amsmath}
\usepackage{float}

\usepackage[utf8]{inputenc}

\graphicspath{{./}{figures/}}

\received{November 26, 2023}
\revised{}
\accepted{July 15, 2024}

\submitjournal{ApJ}

\shorttitle{Realistic Molecular Lines in Cosmological Simulations}
\shortauthors{Garcia et al.}

\bibpunct[, ]{(}{)}{;}{a}{,}{,}
\begin{document}



\title[LIM in Hydro Sims]{{\sc slick}: Modeling a Universe of Molecular Line Luminosities in Hydrodynamical Simulations}

\author[0000-0001-7145-549X]{Karolina Garcia}
\affil{Department of Astronomy, University of Florida, 211 Bryant Space Science Center, Gainesville, FL 32611, USA}
\author[0000-0002-7064-4309]{Desika Narayanan}
\affil{Department of Astronomy, University of Florida, 211 Bryant Space Science Center, Gainesville, FL 32611, USA}
\affil{Cosmic Dawn Center (DAWN), Niels Bohr Institute, University of Copenhagen, Jagtvej 128, K{\o}benhavn N, DK-2200, Denmark}
\author[0000-0003-1151-4659]{Gerg\"o Popping}
\affil{European Southern Observatory, Karl-Schwarzschild-Str. 2, D-85748, Garching, Germany}
\author[0000-0002-3071-3365]{R. Anirudh}
\affil{Indian Institute of Science Education and Research (IISER) Tirupati, Rami Reddy Nagar, Karakambadi Road, Mangalam (P.O.), Tirupati 517 507, India}
\author[0009-0009-3195-3211]{Sagan Sutherland}
\affil{Department of Physics, University of Connecticut, 196 Auditorium Road, U-3046, Storrs, CT 06269, USA}
\author[0000-0002-1173-2579]{Melanie Kaasinen}
\affil{European Southern Observatory, Karl-Schwarzschild-Str. 2, D-85748, Garching, Germany}

\begin{abstract}
We present {\sc slick} (the Scalable Line Intensity Computation Kit), a software package that calculates realistic CO, [\ion{C}{1}], and [\ion{C}{2}] luminosities for clouds and galaxies formed in hydrodynamic simulations. Built on the radiative transfer code {\sc despotic}, {\sc slick} computes the thermal, radiative, and statistical equilibrium in concentric zones of model clouds, based on their physical properties and individual environments. We validate our results applying {\sc slick} to the high-resolution run of the {\sc Simba} simulations, testing the derived luminosities against empirical and theoretical/analytic relations. To simulate the line emission from a universe of emitting clouds, we have incorporated random forest (RF) machine learning (ML) methods into our approach, allowing us to predict cosmologically evolving properties of CO, [\ion{C}{1}] and [\ion{C}{2}] emission from galaxies such as luminosity functions. We tested this model in 100,000 gas particles, and 2,500 galaxies, reaching an average accuracy of $\sim$99.8\% for all lines. Finally, we present the first model light cones created with realistic and ML-predicted CO, [\ion{C}{1}], and [\ion{C}{2}] luminosities in cosmological hydrodynamical simulations, from $z=0$ to $z=10$.
\vspace{1cm}
\end{abstract}



\section{Introduction}
\label{sec:intro}

Star formation occurs within giant molecular clouds (GMCs), which are mostly composed of molecular hydrogen (H$_2$). However, H$_2$ is challenging to observe directly for multiple reasons. First, H$_2$ lacks a permanent dipole moment, resulting in weak radiative transitions. Second, the first excited transition of H$_2$ is at $\sim$300 K, which is difficult to excite in the typical temperatures of $10-100$ K found in cold, molecular gas \citep{fukui_molecular_2010,dobbs_exciting_2013,krumholz_big_2014}. This means that one has to use other trace components of molecular gas to infer the amount of H$_2$ in observed GMCs.

Carbon molecules/atoms such as carbon monoxide, CO, neutral carbon, [\ion{C}{1}], and ionized carbon, [\ion{C}{2}], are the most abundant after H$_2$, and they produce relatively strong and easily observable emission lines, which makes them commonly used tracers of molecular gas \citep[e.g.][]{bolatto_co--h2_2013,vizgan_tracing_2022}, atomic gas, and the star formation rate (SFR) of galaxies. CO has many strong transitions in the submillimeter (sub-mm) range of the spectrum, and has been observed extensively to understand the chemical and structural evolution of galaxies \citep[e.g.][and references therein]{kennicutt_star_2012,carilli_cool_2013,hodge_high-redshift_2020,friascastillo2023}. [\ion{C}{1}] has been detected routinely \citep[e.g.][]{bothwell_alma_2017,popping_alma_2017,friascastillo2024}, as well as ionized carbon [\ion{C}{2}] \citep[e.g.][]{brisbin_strong_2015,capak_galaxies_2015,schaerer_alma_2015,knudsen_c_2016,inoue_detection_2016,rybak2019}, which is a strong coolant in star-forming regions, and a common tracer of SFR. \cite{carilli_cool_2013,olsen_sigame_2017,tacconi_phibss_2018} present compilations of CO, [\ion{C}{1}], and [\ion{C}{2}] observations.

Beyond serving as tracers for the physical conditions of molecular gas, observations of CO, [\ion{C}{1}], and [\ion{C}{2}] also have the potential to constrain cosmological and the evolution of astrophysical parameters in unexplored eras \citep{kovetz_line-intensity_2017,bernal_line-intensity_2022}. Ongoing and upcoming line intensity mapping (LIM) experiments, such as the CO Mapping Array Project \citep[COMAP,][]{cleary_comap_2022}, the CONCERTO project \citep[][]{the_concerto_collaboration_wide_2020}, the EXperiment for Cryogenic Large-Aperture Intensity Mapping \citep[EXCLAIM,][]{pullen_extragalactic_2023}, the Fred Young Submillimeter Telescope \citep[FYST,][]{collaboration_ccat-prime_2022}, the South Pole Telescope Summertime Line Intensity Mapper \citep[SPT-SLIM][]{spt-slim}, the Terahertz Intensity Mapper \citep[TIM,][]{vieira_terahertz_2020}, and the Tomographic Ionized-carbon Mapping Experiment \citep[TIME,][]{holland_time-pilot_2014} will make large-scale intensity maps of these lines from different epochs of the universe to study the cosmic evolution.

Many groups have modeled the physics behind CO, [\ion{C}{1}], and [\ion{C}{2}] emission from galaxies with different levels of complexity in terms of the galaxy formation model, simulation, and luminosity estimation method. Analytical/empirical models \citep[e.g.][]{yue_studying_2019,dizgah_precision_2022,sato-polito2023,zhang_power_2023} provide a straightforward way of making predictions and analyzing the underlying physics of the studied systems, but they typically rely on assumptions based on observations of specific sets of galaxies at low redshifts. Semi-analytic approaches usually combine N-body simulations, galaxy formation models, and radiative transfer methods to infer luminosities \citep[e.g.][]{del_p_lagos_predictions_2012,popping_nature_2014,popping_sub-mm_2016,lagache_cii_2018,popping_art_2019,yang_multitracer_2021}. However, semi-analytic models do not directly simulate the fluid properties in galaxies, instead relying on simplified geometries and analytic prescriptions to describe the physical conditions in star-forming gas.

Cosmological hydrodynamic simulations offer an alternative for computing line emission across cosmic time. 
 Cosmological hydrodynamic simulations fall into two categories: zoom-in simulations, and uniform boxes. 
 Zoom-in simulations follow the line emission properties of individual galaxies, cosmologically, at very high-resolution, though constrain the results to one, or a few individual galaxies \citep[e.g.][]{bisbas_cosmic-ray_2017,pallottini_zooming_2017,pallottini_impact_2017,li_dark_2018,lupi_predicting_2020,pallottini_survey_2022,bisbas_origin_2022,bisbas_pdfchem_2023}. On the other hand, uniform boxes provide large samples of galaxies, though typically have relatively poor resolution compared to zoom-ins; hence sub-resolution modeling is necessary. Beyond this, the cost of running radiative-transfer (RT) equations in all gas particles of such large simulation volumes is significant. For this reason, some groups have relied on empirical relations to make luminosity modeling in large hydrodynamical simulation boxes \citep[e.g.][]{karoumpis_cii_2022,murmu_revisiting_2023}.  Others have combined photoionization models such as {\sc cloudy} with hydrodynamical simulations to compute the cloud luminosities from cosmological simulations \citep{olsen_simulator_2016,olsen_sigame_2017,olsen_sigame_2021,leung_predictions_2020,vizgan_tracing_2022,vizgan_investigating_2022}.

In this paper, we develop a new, flexible, and multi-scale method to capture the microphysics of star-forming clouds in cosmological hydrodynamic simulations. Our framework models line luminosities using a combination of hydrodynamical simulations and molecular gas modeling based on each gas particle's local conditions. We follow the sub-resolution modeling developed in \cite{popping_art_2019}, combined with the chemical network by \cite{gong_simple_2017}, and the radiative transfer code {\sc despotic} \citep{krumholz_despotic_2014} \footnote{https://bitbucket.org/krumholz/despotic}.  We additionally introduce a random forest (RF) machine learning (ML) framework to apply what we learn from modeling a subset of individual clouds/galaxies to entire cosmological volumes in a computationally efficient manner.

This paper is structured as follows. In Section \ref{sec:methods}, we explain the basics of the hydrodynamical simulations that we use here, 
our approach to sub-resolution modeling, and how we compute line luminosities. In Section \ref{sec:slick}, we introduce {\sc slick}. In Section \ref{sec:simba}, we demonstrate applications of {\sc slick} via proof-of-concept applications to the {\sc Simba} cosmological simulation. In Section \ref{sec:discussion}, we discuss the importance of our method, its limitations, and how one could improve upon it. Finally, in Section \ref{sec:conclusion}, we conclude. Throughout this paper, we adopt a flat, cold dark matter cosmology with ${\Omega}_{0} = 0.28$, ${\Omega}_{\Lambda} = 0.72$, $h = H_{0}/(100$ km s$^{-1}$ Mpc$^{-1}) = 0.7$, and $\sigma_8 = 0.812$.

\section{Modeling line luminosities in cosmological simulations}\label{sec:methods}

The main goal of this work is to employ the physical properties of gas particles (or cells) from galaxy evolution simulations in calculating the sub-mm line luminosity of large samples of galaxies. In what follows, we describe our modeling procedure, and in Section \ref{sec:simba} we apply it to the {\sc Simba} cosmological simulation, which is particle-based. However, these methods are general enough that they can be applied to most galaxy evolution models, and adapted for any cosmological hydrodynamical simulation.

First, we describe each gas particle from a hydrodynamical simulation snapshot with a sub-resolution molecular cloud model. The sub-resolution modeling is necessary because current cosmological simulations do not have sufficient resolution to model the internal structures of molecular clouds \citep[though see][]{feldmann_firebox_2023}. Moreover, the majority of cosmological simulations cannot cool to the temperatures relevant for the molecular interstellar medium (ISM), as explained in \cite{crain_hydrodynamical_2023}. This is because the Jeans mass for molecular clouds ($10^4$ $M_\sun$ for a $10$ K cloud) is usually much smaller than the mass-resolution of the simulations ($\sim10^6-10^7$ $M_\sun$). Hence, we need to assume an artificial resolving temperature, and apply sub-resolution modeling. Our sub-resolution model follows the methodology developed by \cite{popping_art_2019} for semi-analytical galaxy formation models. Here we adapt it to be used in hydrodynamical simulation particles.

We then use {\sc despotic} to calculate the statistical, chemical, and thermal equilibria and line luminosities emitting from each sub-resolution molecular cloud. Below, we first describe how we calculate the input properties that the molecular clouds require in {\sc despotic} (density profile, abundances, impinging UV and CR radiation field); then we briefly describe the workings of {\sc despotic}. Table \ref{tab:table} shows a summary of the parameters used and calculated in our model, including information on how/where we get/derive them. For a more detailed description of the sub-resolution model and {\sc despotic} we point the reader to \cite{popping_art_2019} and \cite{krumholz_despotic_2014}, respectively.

\subsection{Sub-resolution modeling} \label{subsec:subgrid}

Each gas particle in a snapshot at redshift $z$, with mass $M_{\rm{C}}$, number density $n_0$, and temperature $T_{\rm{C}}$, is directly extracted from the hydrodynamical simulation. $n_0$ and $T_{\rm{C}}$ will be recalculated later during the radiative-transfer (RT) iterations in our model as in principle these parameters are unrealistic for cold molecular gas. At first, however, particle radii, $R_{\rm{C}}$, are calculated based on their mass and the external pressure, $P_{\rm{ext}}$, acting upon them. Following \cite{faesi_alma_2018,narayanan_physical_2017},
\begin{align}
    \label{eq:radius}
    \frac{R_{\rm{C}}}{\rm{pc}} = \left( \frac{P_{\rm{ext}}/k_{\rm{B}}}{10^4 \; \rm{cm^{-3} K}} \right)^{-1/4} \left( \frac{M_{\rm{C}}}{290 \; \rm{M_{\sun}}} \right)^{1/2},
\end{align}
where $k_{\rm{B}}$ is the Boltzmann constant, and we assume the equation of state,
\begin{align}
    \label{eq:pressure}
    P_{\rm{ext}} = n \; k_{\rm{B}} \; T_{\rm{C}} .
\end{align}

\subsubsection{Density Profile}

We divide each cloud into $16$ different concentric zones, and use a power-law density distribution profile. We have confirmed that the results presented here converge with $16$ zones. \cite{yang_multitracer_2021} found that clouds with a power-law density profile and size determined by the external pressure, enable more accurate constraints on the CO, [\ion{C}{1}], and [\ion{C}{2}] emission when compared to using a constant, logotropic, or Plummer profiles. The density, $n_{\rm{H}}$, is then given by
\begin{align}
    \label{eq:densprofile}
    n_{\rm{H}}(R) = n_0 \left( \frac{R_{\rm{C}}}{R} \right)^{-\alpha} ,
\end{align}
where we adopt $\alpha=2$ \citep{walker_13_1990}.

\subsubsection{Chemical Network}

The hydrogen (H) and carbon (C) chemistry of each zone is computed using the Gong-Ostriker-Wolfire (GOW) chemical network \citep{gong_simple_2017} already implemented in {\sc despotic}. This network was built upon the NL99 network from \cite{nelson_stability_1999} and \cite{glover_approximations_2012}, allowing for predictions in a broader range of physical conditions. It includes eighteen species: H, H$_2$, H$^+$, H$_2^+$, H$_3^+$, He, He$^+$,O,O$^+$,C,C$^+$, CO, HCO$^+$,Si,Si$^+$,e,CH$_x$, and OH$_x$. Twelve of these species are independently calculated, and the remaining six (H, He, C, O, Si and e) are derived by ensuring the conservation of total nuclei and charge. The gas-phase abundance of He nuclei is fixed, while C, O, and Si abundances are taken to be proportional to the gas-phase metallicity relative to the solar neighborhood. The gas-phase metallicity, dust abundance, and grain-assisted reaction rates vary with the gas metallicity. We refer the reader to \cite{gong_simple_2017} for more details.

In this work, we set the abundances for carbon ($x_{\rm{C}}$), oxygen ($x_{\rm{O}}$), and silicon ($x_{\rm{Si}}$) based on the cloud metallicities and the scaling relations between metallicities and these abundances by \cite{draine_physics_2011}. However, it is straightforward to read the abundances directly from the hydrodynamical simulations. We will assess the effectiveness of this approach in a forthcoming paper.

\begin{deluxetable*}{lclc}
\tablecaption{Gas particle (cloud) parameters given as input to radiative transfer code \label{tab:table}}
\tablehead{
\colhead{Parameter} & \colhead{Symbol} & \colhead{Value for each cloud} & \colhead{Units}
}
\startdata
Gas particle's properties \\
\; \; Redshift & z & from hydro sim. &  \\
\; \; Mass & $M_{\rm{C}}$ & from hydro sim. & $M_{\sun}$ \\
\; \; Metallicity & $Z_{\rm{C}}$ & from hydro sim. & $Z_{\sun}$ \\
\; \; Star formation rate & SFR & from hydro sim. & $M_\sun / \rm{yr}$ \\
\; \; Radius & $R_{\rm{C}}$ & Eq. \ref{eq:radius} & pc \\
\; \; External pressure & $P_{\rm{ext}}$ & Eq. \ref{eq:pressure} &  J/cm$^3$ \\
\; \; Density profile & $n_{\rm{H}}$ & Eq. \ref{eq:densprofile} &  1/cm$^3$\\
Radiation field \\
\; \; Ultraviolet radiation field & $\chi$ & Eq. \ref{UV} & Habing \\
\; \; Cosmic ray radiation field & $\rm{\xi}_{\rm{CR}}$ & Eq. \ref{CR} & Habing \\
\; \; Ionization rate & $q$ & $1\times10^{-17} \times G_{\rm{UV}}$ & Habing \\
Dust properties \\
\; \; Dust-to-metal ratio & DTM & Eq. \ref{DTM} &  \\
\; \; Dust abundance relative to solar & $Z_{\rm{d}}$ & $1 \times Z_{\rm{C}} \times \rm{DTM}$ &  \\
\; \; Dust-gas coupling coefficient & $\alpha_{\rm{GD}}$ & $3.2\times10^{-34} \times Z_{\rm{C}} \times \rm{DTM}$ & $\text{erg cm}^3 \text{K}^{-3/2}$ \\
Cross-sections \\
\; \; Cross-section to 10K thermal radiation & $\sigma_{\rm{d,10}}$ & $2\times10^{-25} \times Z_{\rm{C}} \times \rm{DTM}$ & $\text{cm}^2 \text{H}^{-1}$ \\
\; \; Cross-section to 8-13.6 eV photons & $\sigma_{\rm{d,PE}}$ & $1\times10^{-21} \times Z_{\rm{C}} \times \rm{DTM}$ & $\text{cm}^2 \text{H}^{-1}$ \\
\; \; Cross-section to ISRF photons & $\sigma_{\rm{d,ISRF}}$ & $3\times10^{-22} \times Z_{\rm{C}} \times \rm{DTM}$ & $\text{cm}^2 \text{H}^{-1}$ \\
Species abundances \\
\; \; Carbon abundance & $x_{\rm{C}}$ & $1.6\times10^{-4} \times Z_{\rm{C}}$ &  \\
\; \; Oxygen abundance & $x_{\rm{O}}$ & $3.2\times10^{-4} \times Z_{\rm{C}}$ &  \\
\; \; Silicon abundance & $x_{\rm{Si}}$ & $1.7\times10^{-6} \times Z_{\rm{C}}$ &  \\
\enddata
\end{deluxetable*}

\begin{figure*}%
    \centering
    \includegraphics[width=14.5cm]{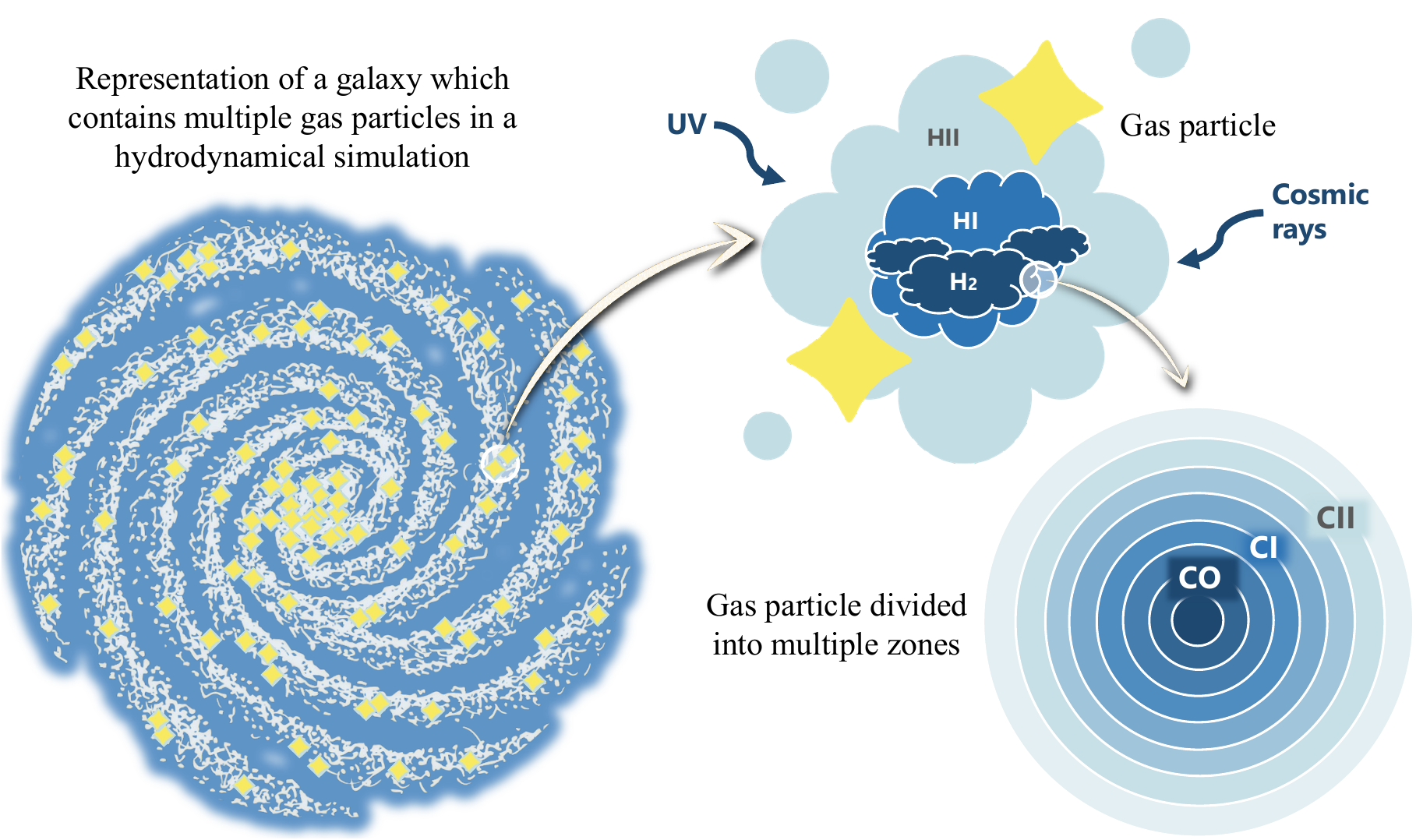}%
    \caption{\textbf{Representation of our model.} Gas particle properties and galaxy catalogs come from the hydrodynamical simulations. Each galaxy may contain a few to thousands of gas particles. Gas particle sizes are set as a function of the molecular cloud mass and the external pressure acting on the molecular clouds. They are illuminated by a far-UV radiation field and cosmic rays, based on the gas particles around them. Each gas particle has a radial density profile, and they are radially divided into concentric zones which must achieve chemical, thermal, and statistical equilibrium within our model.}%
	\label{fig:slick1}%
\end{figure*}

\subsubsection{Radiation Field}

The radiation field impinging upon molecular clouds, especially the ultraviolet (UV) and cosmic ray (CR) fields, affect the chemistry within the clouds. \cite{popping_art_2019} showed that [\ion{C}{2}] luminosities are more sensitive to the radiation field model than CO and [\ion{C}{1}]. This happens because a stronger radiation field ionizes more carbon on the outskirts of the cloud, but it also  increases the temperature and optical depth, consequently not affecting much CO and [\ion{C}{1}] luminosities. Whereas [\ion{C}{2}] only gets stronger with higher temperatures and ionization rates.

There are different approaches to model the strength of UV and CR fields. For instance, \cite{narayanan_physical_2017} scale the strengths with the integrated star-formation rate (SFR) of galaxies, and \cite{popping_art_2019} scale them with local SFR surface densities. Here we follow \cite{del_p_lagos_predictions_2012}, and compute a local SFR surface density by summing up the SFRs of the 64 nearest gas particles (the same number of particles that {\sc Simba} smoothes its mass content over) and dividing it by a cross-sectional area containing these clouds. To account for attenuation, we compute the dust mass ($M_{\rm{dust}}$) density within this cross-sectional area and convert it to a UV optical depth,
\begin{align}
    \label{tau}
    \tau_{\rm{UV}} = \kappa_{\rm{abs}} \times R_{\rm{\sigma}} \times \frac{3 M_{\rm{dust}}}{4\pi R^3_{\rm{\sigma}}} ,
\end{align}
where $\kappa_{\rm{abs}}$ is the mass attenuation coefficient/absorption cross-section per mass of dust, and $R_{\sigma}$ is the radius of the 64-particle sphere. We use $\kappa_{\rm{abs}} = 1.078\times10^5 \; {\rm{cm}}^2 {\rm{g}}^{-1}$, which is the median value of the mass attenuation coefficients within the Habing limit of 91.2 nm to 111.0 nm (range at which CO and H$_2$ photoexcitation and/or photoionization usually occur), assuming a Milky Way (MW) relative variability of $R_v = 3.1$ \footnote{\label{drainefootnote}\url{https://www.astro.princeton.edu/~draine/dust/extcurvs/kext\_albedo_WD_MW_3.1_60_D03.all}}.

The {\sc Simba} cloud neighborhood’s UV transmission probability is then \citep{del_p_lagos_predictions_2012}:
\begin{align}
    \label{beta}
    \beta_{\rm{UV}} = \frac{1-e^{-\tau_{\rm{UV}}}}{\tau_{\rm{UV}}} .
\end{align}
This is further normalized by the Solar neighborhood value, which we calculate using a MW gas surface density of $10 \; M_{\sun}{\rm{pc}}^{-2}$ \citep{chang_history_2002}, and a gas-to-dust ratio of 165 $^{\ref{drainefootnote}}$, which gives ${\beta}_{\rm{UV,MW}}\approx0.5$. So the cloud's UV radiation field is finally
\begin{align}
    \label{UV}
    \chi = \frac{\Sigma_{\rm{SFR,C}}}{\Sigma_{\rm{SFR,MW}}} \times \frac{\beta_{\rm{UV,C}}}{\beta_{\rm{UV,MW}}} ,
\end{align}
where ${\Sigma}_{\rm{SFR,C}}$ is the SFR surface density of the cloud, and ${\Sigma}_{\rm{SFR,MW}}$ is the SFR surface density of the Solar neighborhood, equal to $790 \; M_{\sun} {\rm{Myr}^{-1}} {\rm{kpc}^{-2}}$ \citep{bonatto_constraining_2011}.

The CR field is then scaled with the radiation field as
\begin{align}
    \label{CR}
    \xi_{\rm{CR,C}} = \zeta_{-16} \times \xi_{\rm{CR,MW}} \times \chi ,
\end{align}
where $\zeta_{-16} = 0.1$ is the CR ionization rate, and $\xi_{\rm{CR,MW}} = 10^{-16}$ s$^{-1}$ is the CR field in the Solar neighborhood, both following \cite{narayanan_physical_2017}.

\subsubsection{Dust Properties}

Dust plays an important role in the physics of the ISM; dust grains act as catalysts for the formation of molecules and enable the gas to cool more effectively. Here we highlight the dust property assumptions used in our model.

The cross-sections ($\sigma_{\rm{d,10}}$, $\sigma_{\rm{d,PE}}$, $\sigma_{\rm{d,ISRF}}$), dust abundance ($Z_{\rm{d}}$), and dust-gas coupling coefficient ($\alpha_{\rm{GD}}$) of each cloud are set to scale with the cloud's dust-to-metal ratio and metallicity, as shown in Table \ref{tab:table}. The dust-to-metal ratio (DTM) follows \cite{li_dust--gas_2019} as
\begin{align}
    \label{DTM}
    \rm{DTM} = \frac{\rm{DTG}}{0.44 \times Z_{\rm{G}}}  ,
\end{align}
where
\begin{align}
    \label{DTG}
    \log{(\rm{DTG})} = 2.445 \times \log{\frac{Z_{\rm{G}}}{0.0134}-2.029} ,
\end{align}
DTG is the dust-to-gas ratio,  and $Z_{\rm{G}}$ is the cloud's host galaxy metallicity.

\subsection{Computing Line Emission}

We use the python package {\sc despotic} \citep{krumholz_despotic_2014} to calculate the spectral line luminosities of the clouds. {\sc despotic} takes the set of cloud properties described in Subsection \ref{subsec:subgrid} and Table \ref{tab:table}, and simultaneously iterates on the thermal, chemical, and radiative equilibrium equations until convergence is reached in each of the concentric zones.  We illustrate this schematically in Fig.~\ref{fig:slick1} and describe the luminosity calculation process in more detail below.

First, {\sc despotic} assumes statistical equilibrium of the various level populations. The level populations are determined by the balance between the rate of transitions of each species into and out of each level, according to
\begin{align}
    \sum_{j} &f_j \left[ q_{ji} + (1 + n_{\gamma,ji}) A_{ji} + \frac{g_i}{g_j}n_{\gamma,ij} A_{ij} \right] \\
    &= f_i \sum_k \left[ q_{ik} + (1 + n_{\gamma,ik}) A_{ik} + \frac{g_k}{g_i}n_{\gamma,ki} A_{ki} \right] ,
\end{align}
where $f_i$ ($f_j$) is the fraction of atoms/molecules in the state $i$ ($j$); $A_{ij}$ ($A_{ji}$) is the Einstein coefficient for absorption (emission); and $g_i$ ($g_j$) represents the degeneracy of the state $i$ ($j$). The parameters
\begin{align}
    n_{\gamma,ij} &= \frac{1}{\exp{(\Delta E_{ij}/k_{\rm{B}} T_{\mathrm{CMB}})}-1} ,\\
    q_{ij} &= f_{\mathrm{cl}} \; n_{\mathrm{H}} \sum_p x_p \; k_{p,ij}
\end{align}
are the photon occupation number at the frequency of the line connecting states $i$ and $j$, and the rate of collisional transitions between states $i$ and $j$ summed over all collision partners $p$, respectively. $\Delta E_{ij}$ is the absolute value of the energy emitted/absorbed, $f_{\rm{cl}}$ is a clumping factor, $n_{\rm{H}}$ is the density of the cloud zone, $x_p$ is the abundance of species $p$, and $k_p$ is the rate coefficient for collisional transitions.

Thermal equilibrium is reached by setting the rate of change of the gas energy per H nucleus equal to zero. The energy rate is
\begin{equation}
    \frac{de_{\mathrm{gas}}}{dt} = \Gamma_{\mathrm{ion}} + \Gamma_{\mathrm{PE}} + \Gamma_{\mathrm{grav}} - \Lambda_{\mathrm{line}} + \Psi_{\mathrm{gd}} ,
\end{equation}
where $\Gamma_{\mathrm{ion}}$, $\Gamma_{\mathrm{PE}}$, and $\Gamma_{\mathrm{grav}}$ are the rates of ionization, photoelectric, and gravitational heating per H nucleus, $\Lambda_{\mathrm{line}}$ is the rate of line cooling per H nucleus, and $\Psi_{\mathrm{gd}}$ is the rate of dust-gas energy exchange per H nucleus. {\sc despotic} then solves for equilibrium gas temperature by setting $de_{\mathrm{gas}}/dt = 0$.

Finally, we use {\sc despotic} to integrate the chemical evolution equations,
\begin{equation}
    \frac{d\vec{x}}{dt} = f(\vec{x},n_{\mathrm{H}},N_{\mathrm{H}},T_{\mathrm{g}},\Xi,...) ,
\end{equation}
where $\vec{x}$ is the vector of fractional abundances for the various species taken into account. The reaction rates $d\vec{x}/dt$ can be a function of these abundances, the number density of H, $n_{\mathrm{H}}$, the column density, $N_{\mathrm{H}}$, the gas temperature, $T_{\mathrm{g}}$, the ionization rate, $\Xi$, or any of the other parameters {\sc despotic} uses to define a cloud. This series of equations is iterated over many times until convergence is reached inside of each zone in each simulation cloud. Finally, the zones are combined to give the full resulting line emission of different species from the cloud.

\section{{\sc slick}: Scalable Line Intensity Computation Kit} \label{sec:slick}

\begin{figure*}%
    \centering
    \includegraphics[width=14.5cm]{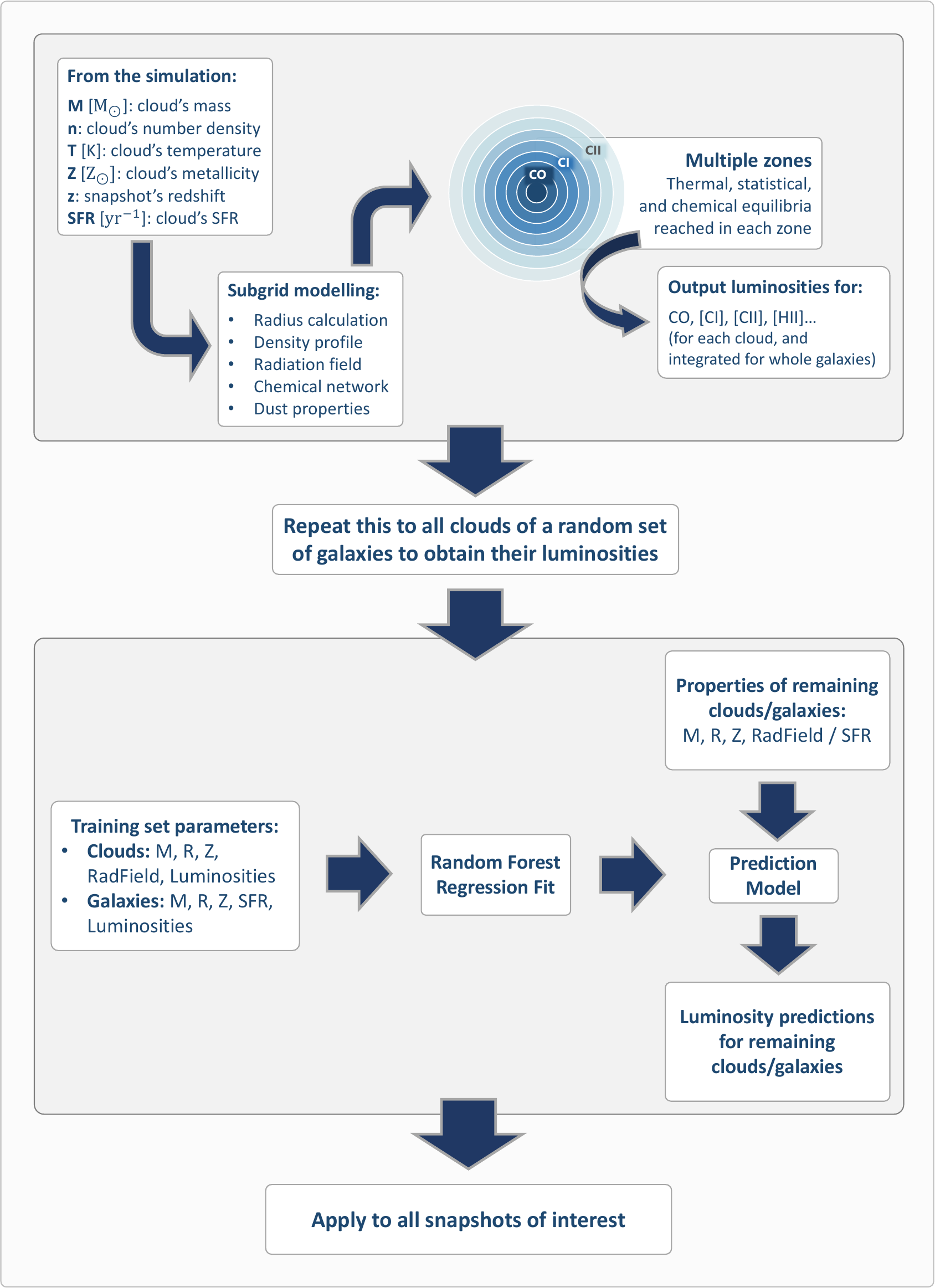}%
    \caption{\textbf{Schematic representation of the steps involved in our exact luminosity calculation + ML modeling.} The gas particle mass, number density, temperature, metallicity, redshift, and SFR are extracted directly from the simulation. Cloud radii are calculated using the external pressure, which we calculate based on the cloud number density and mass. The radiation field depends on the neighboring gas particles. For sub-resolution modeling, we assume the initial densities and fractions, a density profile that divides each gas particle in many zones, a chemical network, and the dust properties. All of this information is fed into {\sc despotic}, which keeps computing abundances until each zone reach thermal, statistical and chemical equilibria, and outputs the luminosities of interest. This is repeated for all gas particles of a given galaxy. One can choose to speed up the process by only running this for certain clouds/galaxies, and using the ML framework to predict luminosities for new clouds/galaxies. A Random Forest model takes clouds' mass, radius, metallicity and radiation field as input parameters and outputs new predicted luminosities. For galaxies, the input parameters are galaxy mass, radius, metallicity, and SFR. Finally, one can apply this to all snapshots (redshifts) of interest to obtain a light cone.}%
	\label{fig:slick2}%
\end{figure*}

Obtaining realistic light cones would, at first glance, require that all the processes explained in Section \ref{sec:methods} are applied to most (if not all) gas particles in full hydrodynamical simulation snapshots. However, this process of iterating over thermal, chemical, and radiative equilibrium networks for hundreds of millions of gas particles, each divided into many zones, is computationally expensive, and generally unfeasible in reasonable timescales.

In light of this, we here present {\sc slick} \footnote{https://karolinagarcia.github.io/slick}: a framework that makes use of the exact cloud-by-cloud luminosity calculation methods described in Section \ref{sec:methods}, and offer an additional solution for optimization and scalability of these computations through ML. {\sc slick} computes both cloud and integrated galaxy realistic luminosities for different molecular lines in snapshots of cosmological hydrodynamical simulations such as {\sc Simba} \citep{dave_simba_2019} and IllustrisTNG \citep{nelson_illustris_2015,pillepich_first_2018,donnari_star_2019}.

In detail, {\sc slick} provides the option to apply the sub-resolution modeling and the radiative transfer equations to all gas particles of a snapshot; or to run them only on a random subset of gas particles, using an RF model to predict luminosities for the remaining ones. {\sc slick}'s workflow is summarized in Fig.~\ref{fig:slick2}. The RF model predicts gas particle luminosities based on their mass, radius, metallicity, and radiation field. We found RF to be the simplest model that can handle complex, non-linear relationships within our data. Besides optimizing the time spent on the luminosity computation within each simulation box, our ML framework also allows for calculation of entire new snapshots if trained on a box with the same resolution and cloud assumptions. For example, one can train the RF on redshifts $z=1$ and $z=2$ and predict the results for $z=1.5$ if the same simulation suite is used for both.

We additionally study the ML mapping of galaxy properties to modeled line emission strengths. In other words, besides gas particle luminosities, our RF model can also predict the integrated line luminosities of galaxies based on their mass, radius, metallicity, SFR, and redshift. This allows us to apply our simulation results to simulations of varying resolutions, and expand the box sizes in future works in this project. We test {\sc slick} on the {\sc Simba} simulations, showing its application to {\sc Simba} 25 Mpc/h boxes, along with validation plots, luminosity functions and light cones in Section \ref{sec:simba}.

\section{{\sc slick} on {\sc Simba}} \label{sec:simba}

{\sc Simba}\footnote{https://simba.roe.ac.uk} \citep{dave_simba_2019} is a suite of galaxy formation simulations developed on the meshless finite mass hydrodynamics of {\sc Gizmo} \citep{hopkins_new_2015}, a code originated from {\sc Gadget-3} \citep{springel_aquarius_2008}. Compared to its predecessor, {\sc Mufasa}, {\sc Simba} has more realistic sub-resolution prescriptions for star formation, AGN feedback, and black hole growth \citep{dave_mufasa_2016}. {\sc Simba}'s outputs consist of snapshots and galaxy catalogs for $151$ redshifts from $z=20$ to $z=0$, for three separate volumes of $25$, $50$, and $100$ Mpc/h per side.

To test our model, we use the highest resolution {\sc Simba} simulations, which are $25$ Mpc/h-length cubes containing $512^3$ particles, with a baryon mass resolution of $1.4 \times 10^6$ $M_\sun$. The cross-matching between galaxies and halo physical properties is found on the {\sc caesar}\footnote{https://github.com/dnarayanan/caesar} catalogs \citep{thompson_pygadgetreader_2014}, which identify galaxies using 6-D friends-of-friends, 
compute galaxy physical properties, and outputs {\sc Simba} galaxies’ information in a HDF5 catalog. This means that {\sc Simba} and {\sc caesar} together provide a mock catalog of clouds, their host galaxies, and their physical properties, such as: mass, number density, temperature, metallicity, etc.

\subsection{ML Model}

\begin{figure*}[htb!]
\centering
    \includegraphics[width=0.47\linewidth]{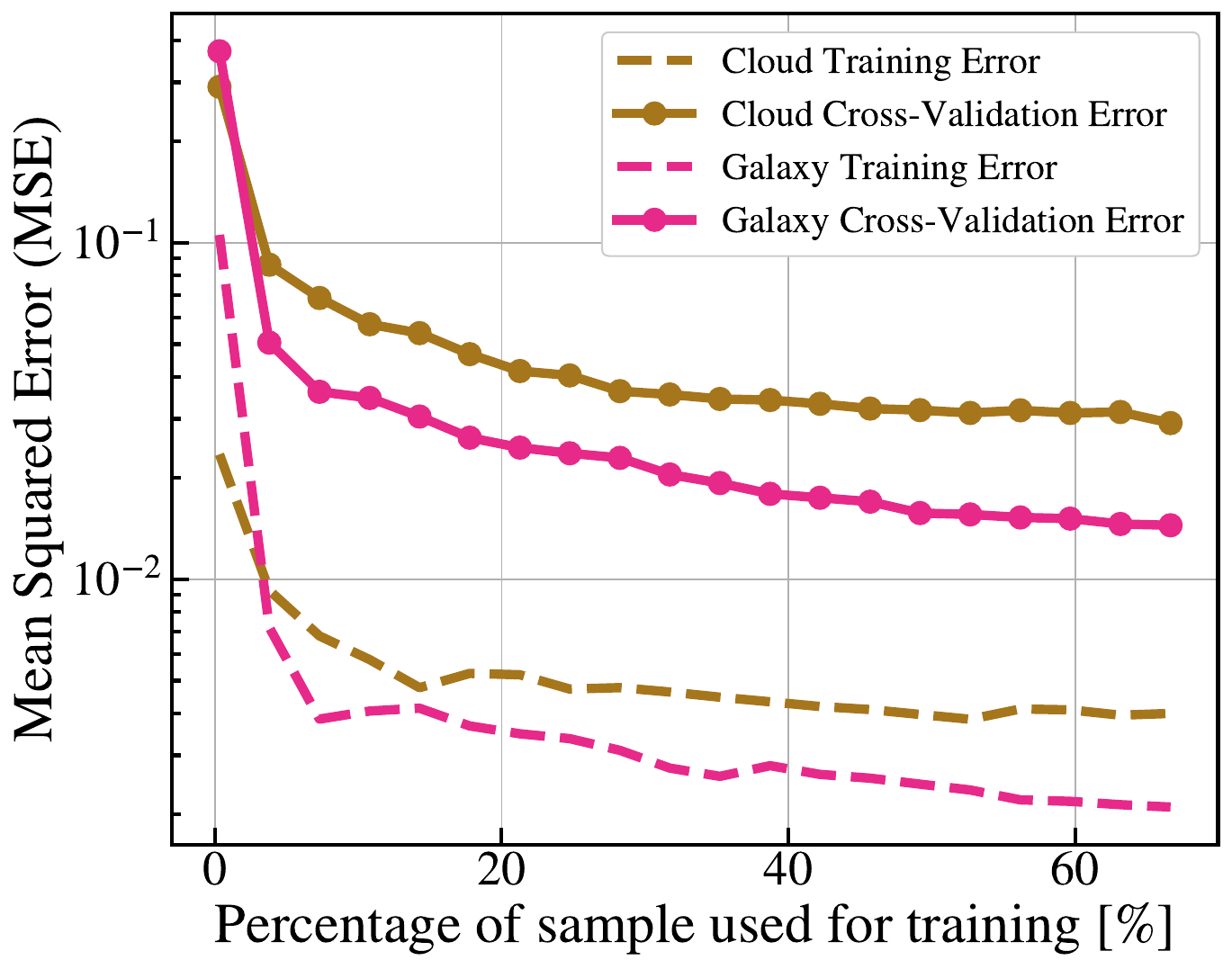}\hfil
    \includegraphics[width=0.48\linewidth]{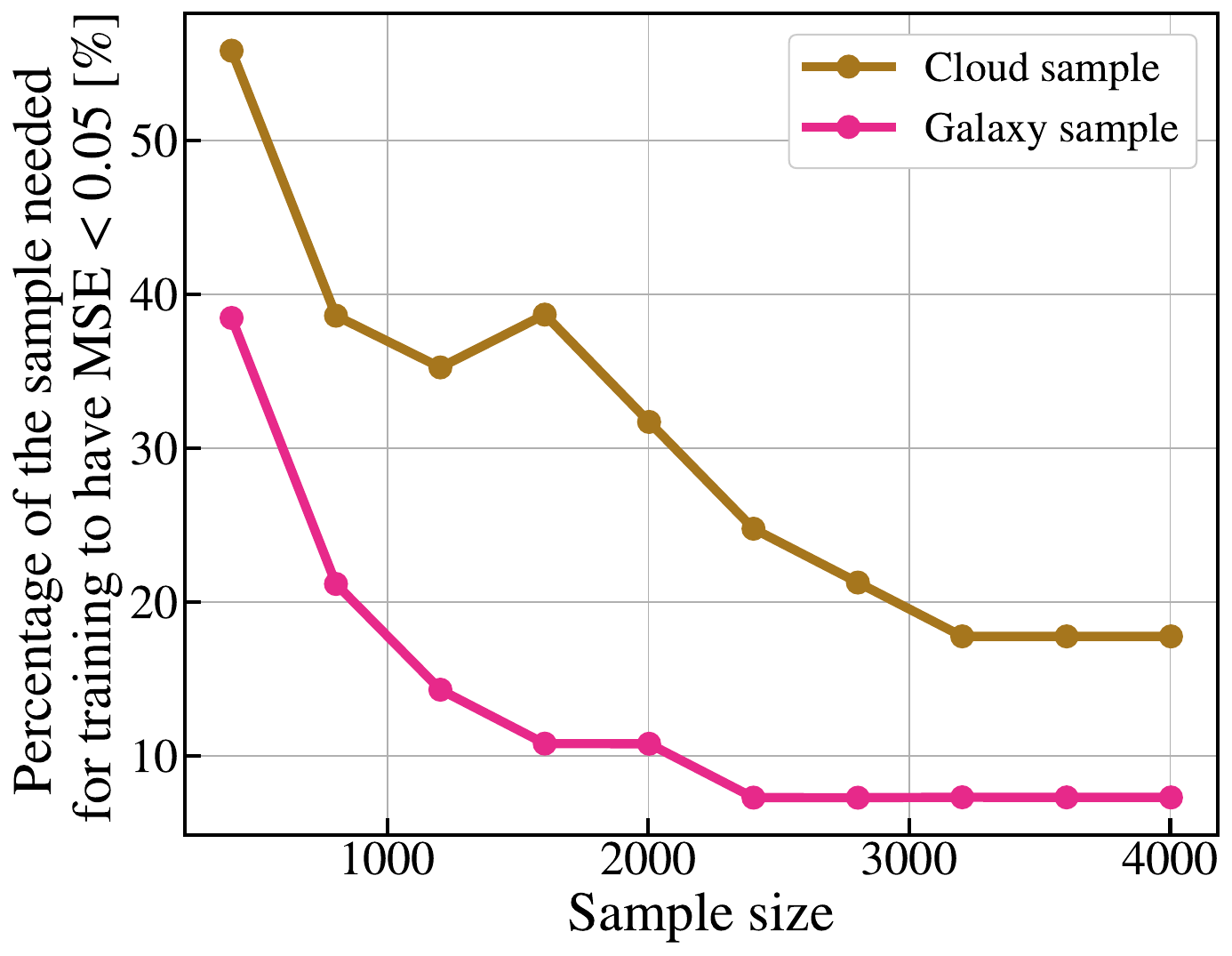}\hfil
\caption{\textbf{We use a random selection of 4,000 clouds and 4,000 galaxies from the $z=0, 1, 2,8$, and $10$ snapshots to investigate the efficacy of our RF model and what fraction of our sample we need to train to predict for the entire dataset.} [Left] Prediction mean squared error for different fractions of the same dataset to train. The training error is the model's performance on the same dataset it learns from, and the cross-validation error measures the model’s ability to generalize to new, unseen data. [Right] Fraction of the sample that needs to be trained to achieve a less-than-0.05 cross-validation error, for different sample sizes. We find that 7\% of the galaxy sample is needed to achieve an MSE$\leq0.05$, and 17\% of the cloud sample.}
\label{fig:MSE}
\end{figure*}

As explained in Section \ref{sec:slick}, our framework provides an option to predict cloud/galaxy luminosities based on an RF ML model. The motivation is twofold: not having to run radiative transfer equations in all clouds of the cosmological simulations, and to be able to train our model on galaxies from high resolution runs and predict for lower resolution simulations in upcoming projects. In what follows, we explore capacities and limitations of this model, and provide information on how we picked our training sample sizes.

First, we investigate how large of a sample we need to train our model on to obtain reliable results. Using a random selection of 4,000 clouds and 4,000 galaxies from the $z=0, 1, 2,8$, and $10$ snapshots, we show in the left plot of Fig. \ref{fig:MSE} how the prediction mean squared error (MSE) changes when using different fractions of the same dataset to train. We present both the training error (when predictions are made for the same subset they got trained on) and the cross-validation error (when predictions are made for the subsample that was not used for training). As the training error represents the model's performance on the same dataset it learns from, it typically decreases as more data is provided due to better learning and fitting capabilities. In our plot, it starts around 0.01 for both clouds and galaxies, and reaches $\sim 0.02 - 0.04$ around a sample percentage of $40\%$, where it stays mostly constant. The cross-validation error measures the model’s ability to generalize to new, unseen data, not used during the training process. This error also starts high when the model is trained on few data points, and decreases as more data is used, which helps the model generalize better. In our plot we see that for training samples of less than $10 \%$, our model overfits, since it shows a very small training error, but very high cross-validation error. As the training sample size increases, the cross-validation error drops and stabilizes, meaning that we have reached a good level of generalization.

In the right plot of Fig. \ref{fig:MSE} we check, for different sizes of cloud/galaxy data, what fraction needs to be trained to achieve a less-than-0.05 cross-validation error. For example, for a dataset containing 2,000 clouds, we need to train on at least $\sim 30\%$ of it to achieve a cross-validation error of less than 0.05. For a dataset of 3,000 and above a constant percentage of $\sim 17\%$ is needed. In the case of galaxies, a dataset of 2,500 and above needs training on $\sim 7\%$ of it. The 25 Mpc/h $z=0$ box contains 2,261 galaxies, so running full-RT {\sc slick} on $\sim 160$ galaxies and training the ML model on them would make reliable predictions for the remaining 2,171.

\begin{figure*}
\centering
    \includegraphics[width=0.28\linewidth]{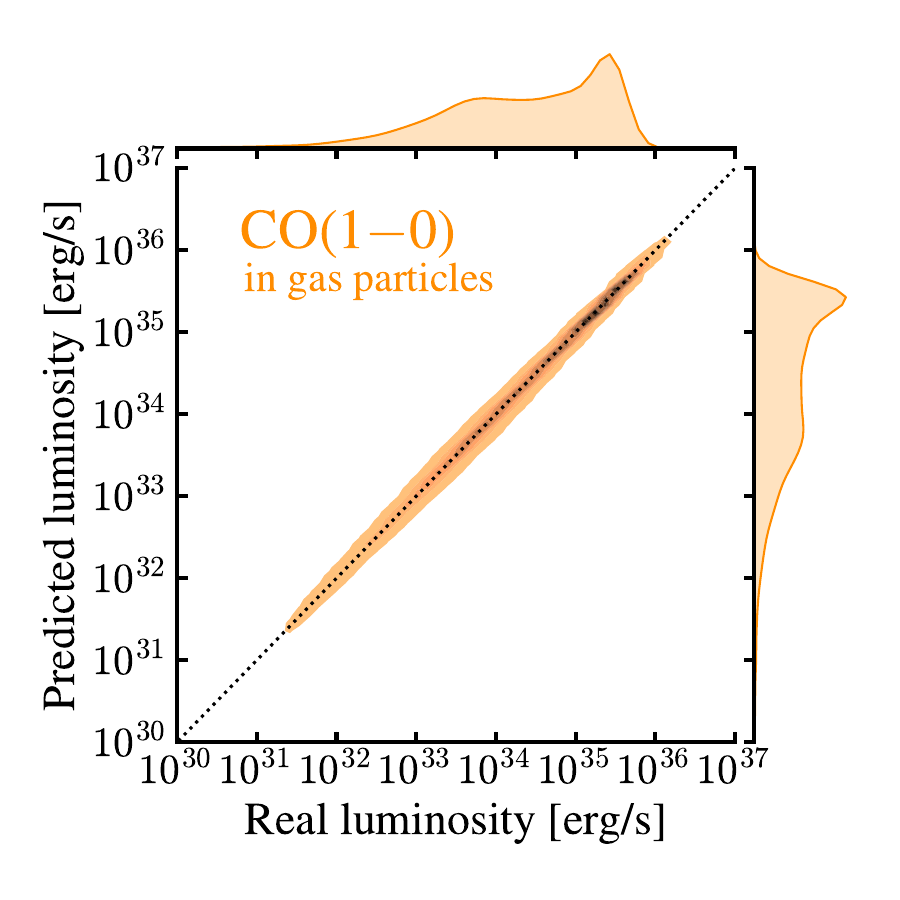}\hfil
    \includegraphics[width=0.28\linewidth]{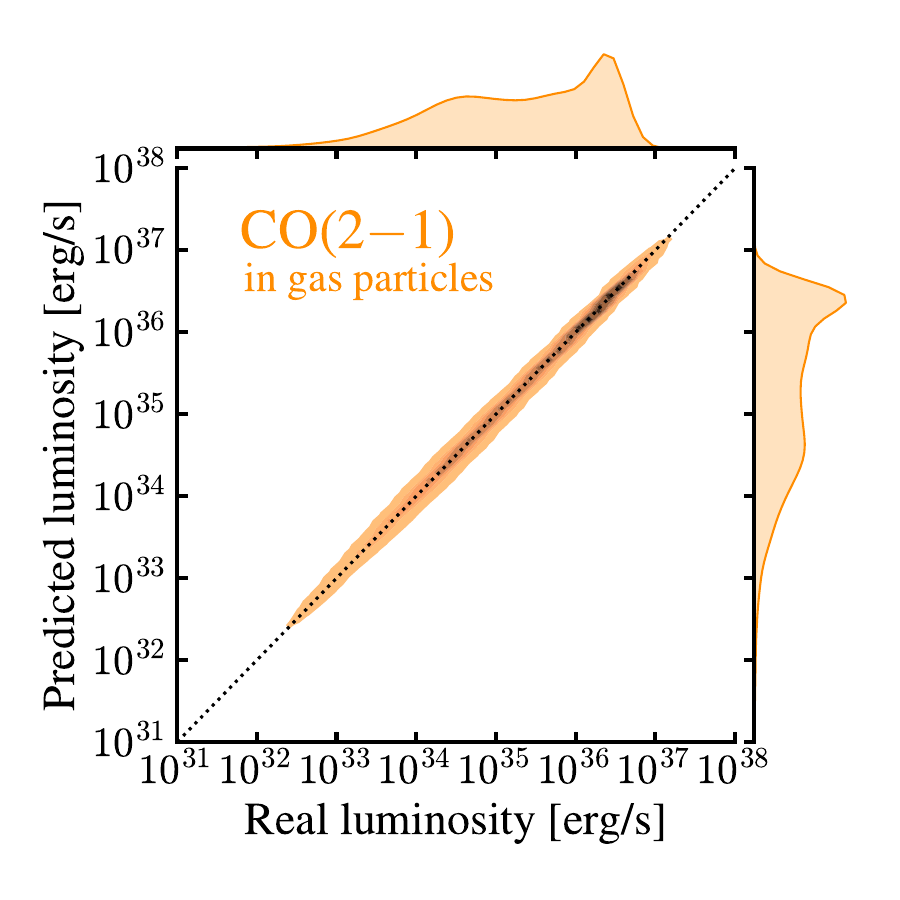}\hfil
    \includegraphics[width=0.28\linewidth]{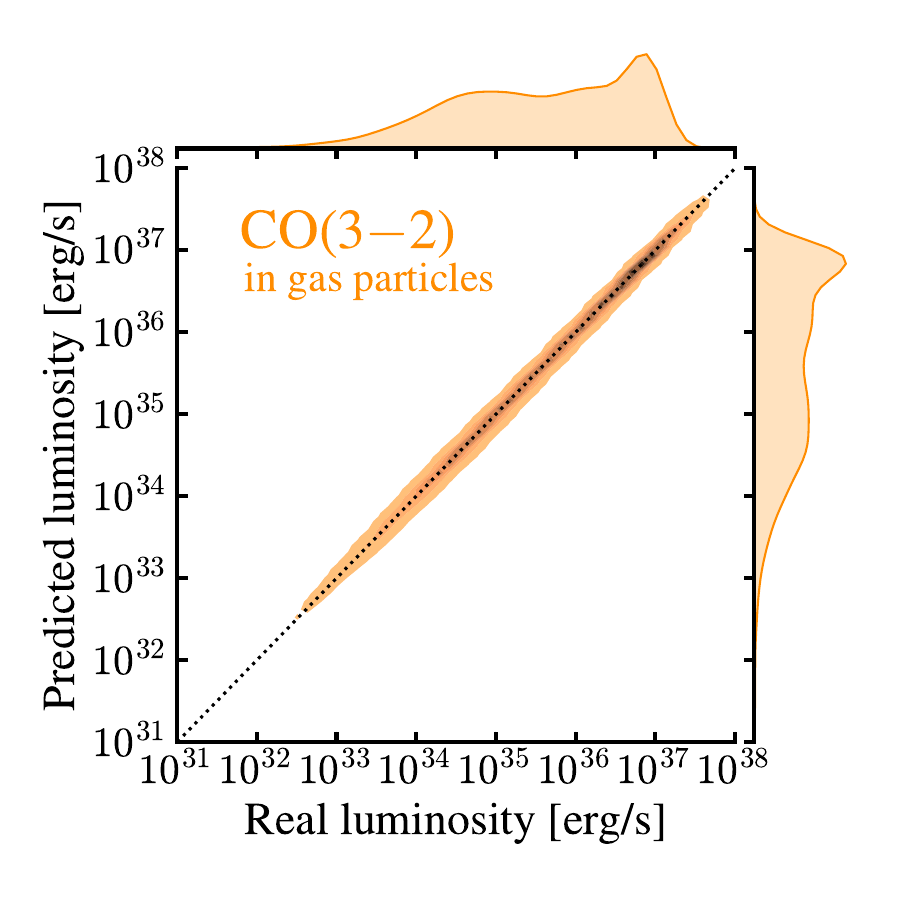}\hfil
    \includegraphics[width=0.28\linewidth]{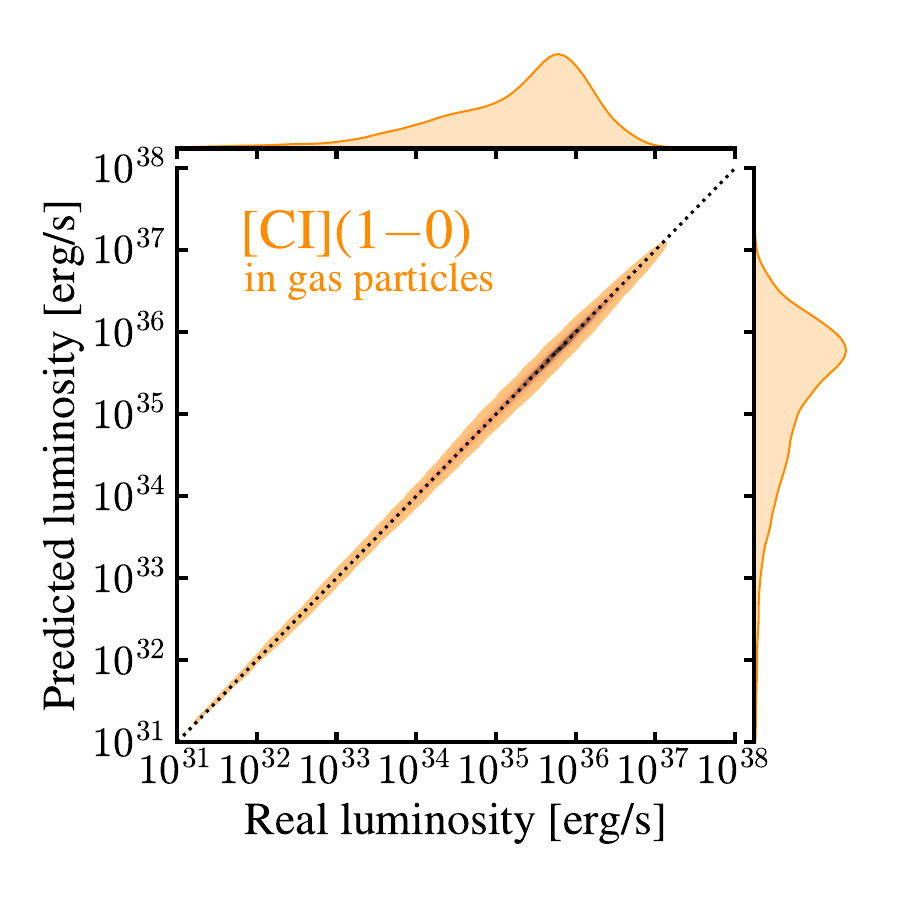}\hfil
    \includegraphics[width=0.28\linewidth]{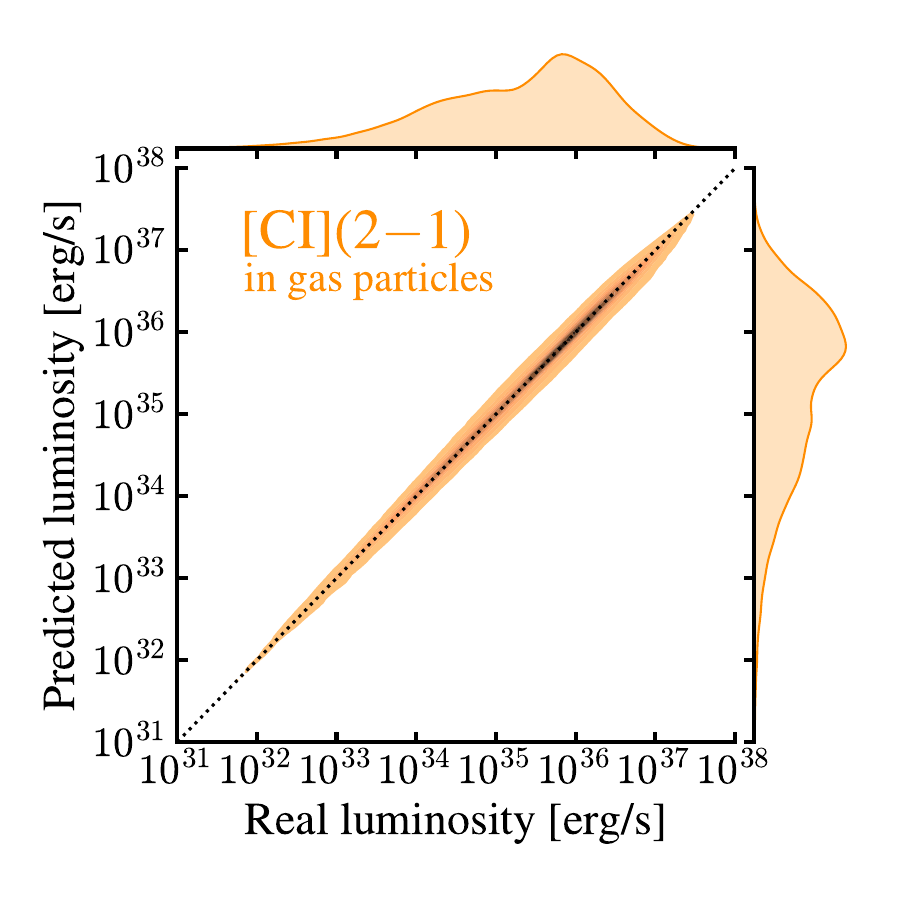}\hfil
    \includegraphics[width=0.28\linewidth]{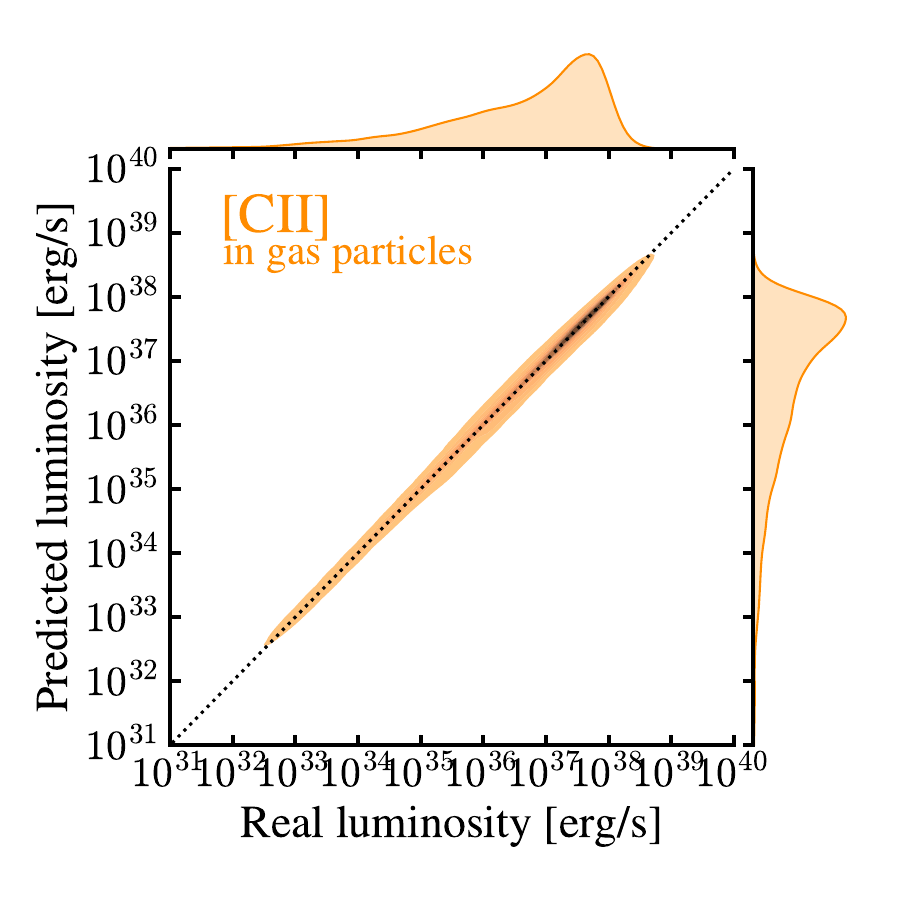}\hfil
\caption{\textbf{Random Forest Regression model predicts {\sc slick}'s cloud luminosities with an accuracy of 99.8\%.} We show kernel density estimation (KDE) plots comparing 80,000 real gas particle luminosities to their predictions using our RF-based ML model. Each plot is for the indicated lines: CO(1-0), CO(2-1), CO(3-2), [\ion{C}{1}](1-0), [\ion{C}{1}](2-1), and [\ion{C}{2}].}
\label{fig:ml_clouds}
\end{figure*}

\begin{figure*}
\centering
    \includegraphics[width=0.26\linewidth]{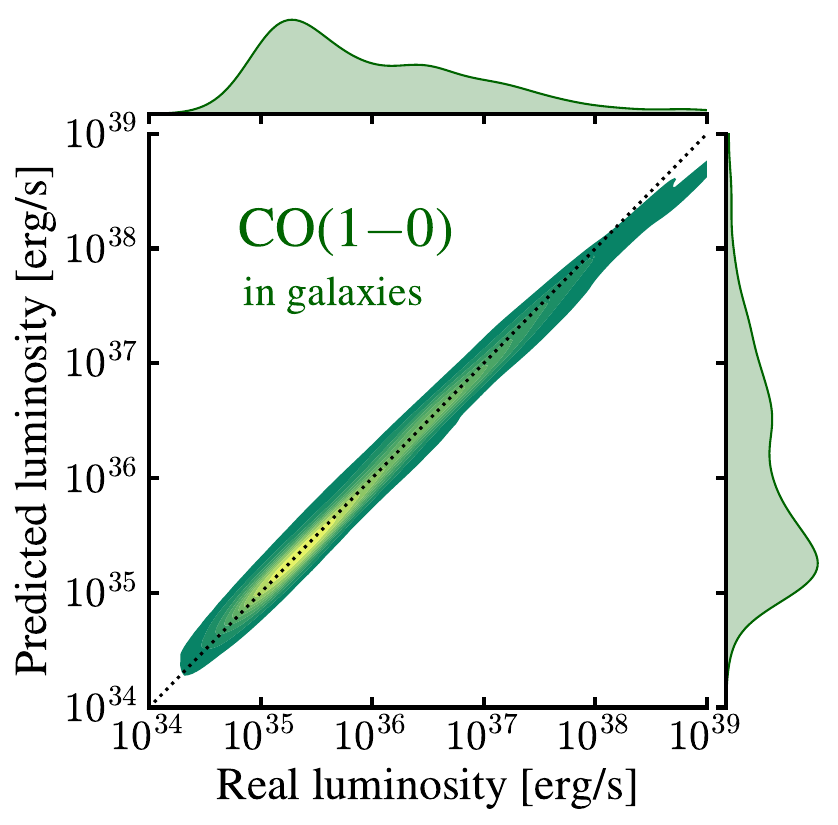}\hfil
    \includegraphics[width=0.26\linewidth]{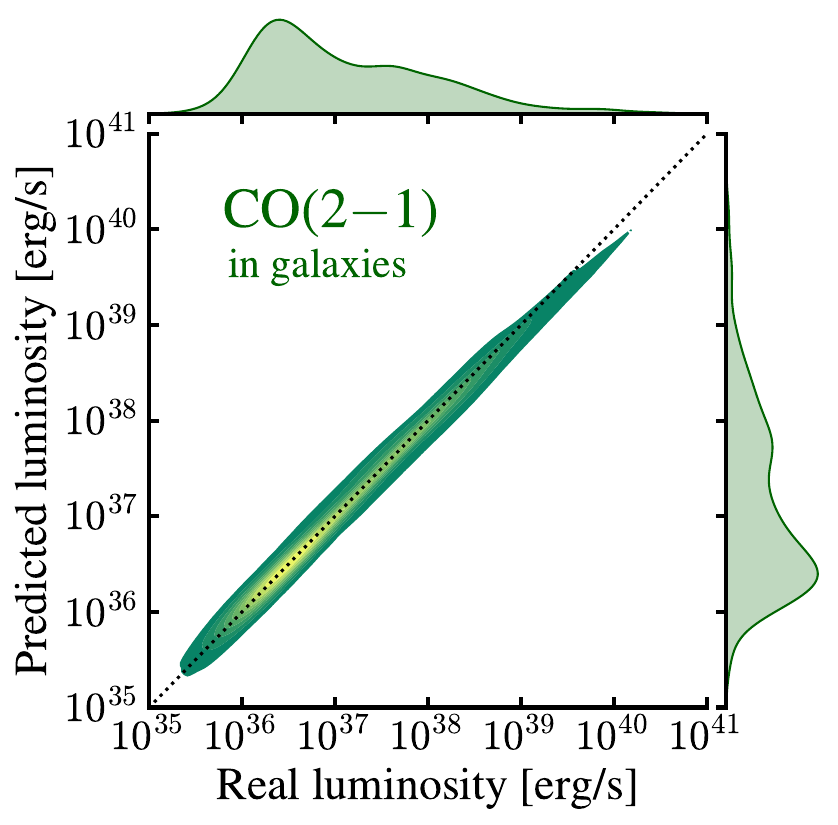}\hfil
    \includegraphics[width=0.26\linewidth]{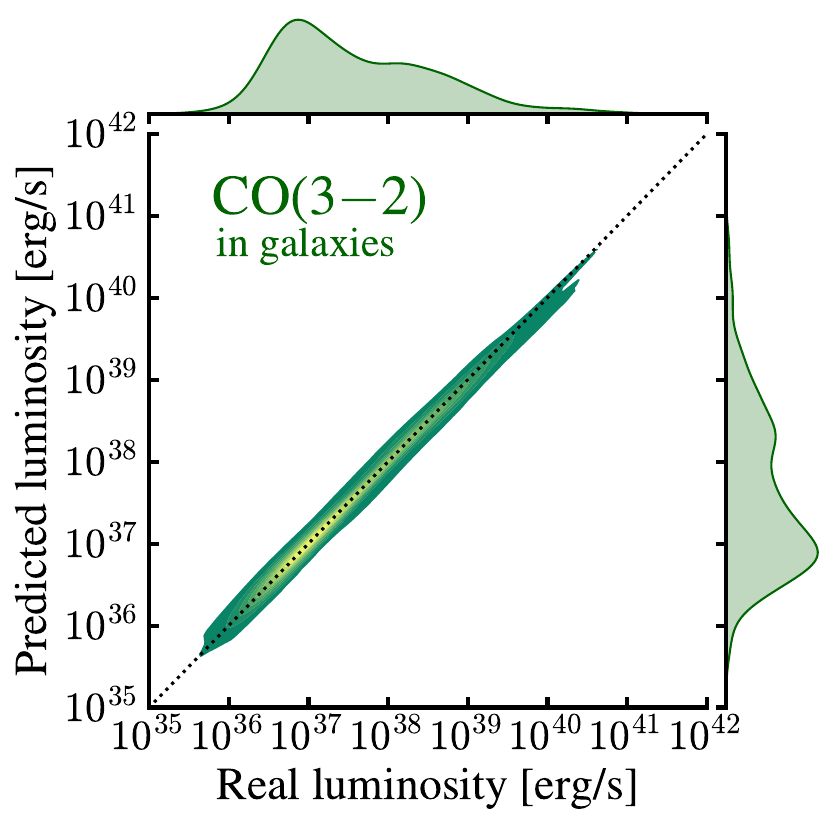}\hfil
    \includegraphics[width=0.26\linewidth]{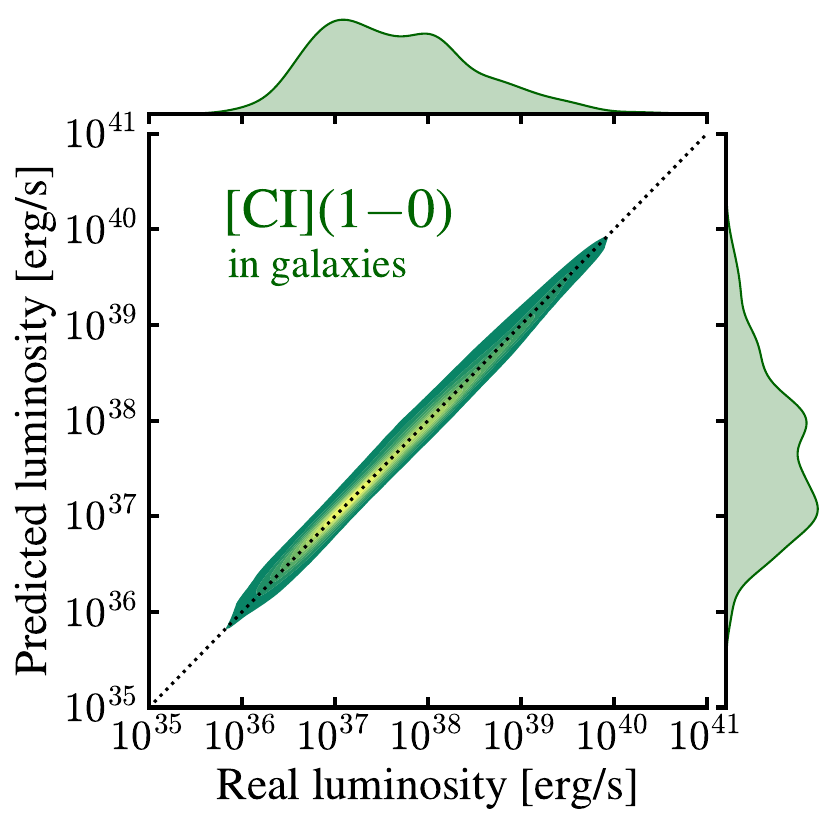}\hfil
    \includegraphics[width=0.26\linewidth]{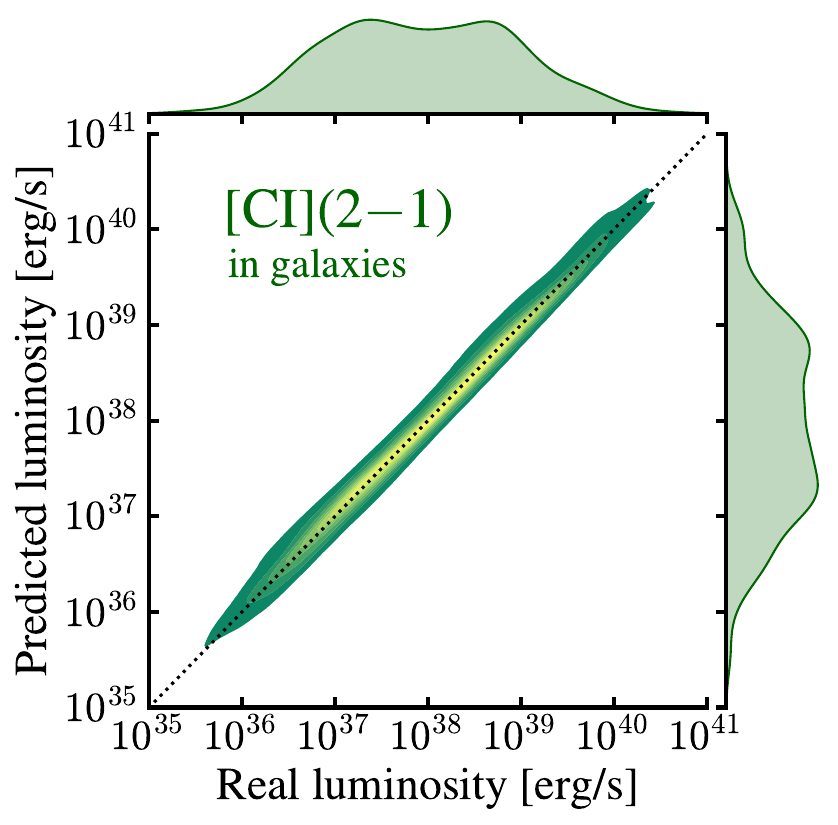}\hfil
    \includegraphics[width=0.26\linewidth]{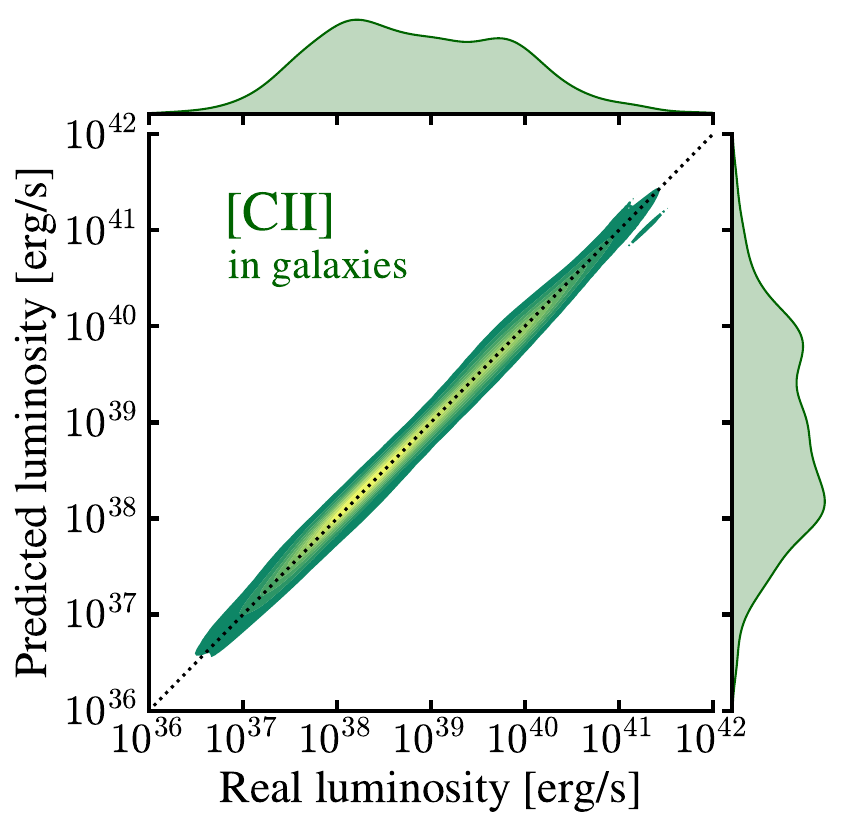}\hfil
\caption{\textbf{Random Forest Regression model predicts {\sc slick}'s galaxy luminosities with an accuracy of 99.8\%.} We show kernel density estimation (KDE) plots to compare 3,200 real galaxy luminosities to their predictions using our RF-based ML model. Each plot is for the indicated lines: CO(1-0), CO(2-1), CO(3-2), [\ion{C}{1}](1-0), [\ion{C}{1}](2-1), and [\ion{C}{2}].}
\label{fig:ml_gals}
\end{figure*}

Another way of visualizing the performance of our model is by showing a direct comparison between real and predicted values for different luminosities of our samples. Here we decide to extend our cloud sample to 100,000 since we have a very large sample of clouds with exact luminosities computed, and keep using a sample of 4,000 galaxies. Based on the analysis provided by Fig. \ref{fig:MSE}, we see that $20\%$ of the data is a good training sample for both cloud and galaxy datasets. So we use 20,000 clouds to train and predict for the remaining 80,000; and 800 galaxies to train and predict for the remaining 3,200. Figures \ref{fig:ml_clouds} and \ref{fig:ml_gals} show the performance in such predictions.

\begin{figure*}[htb!]
\centering
    \includegraphics[width=0.48\linewidth]{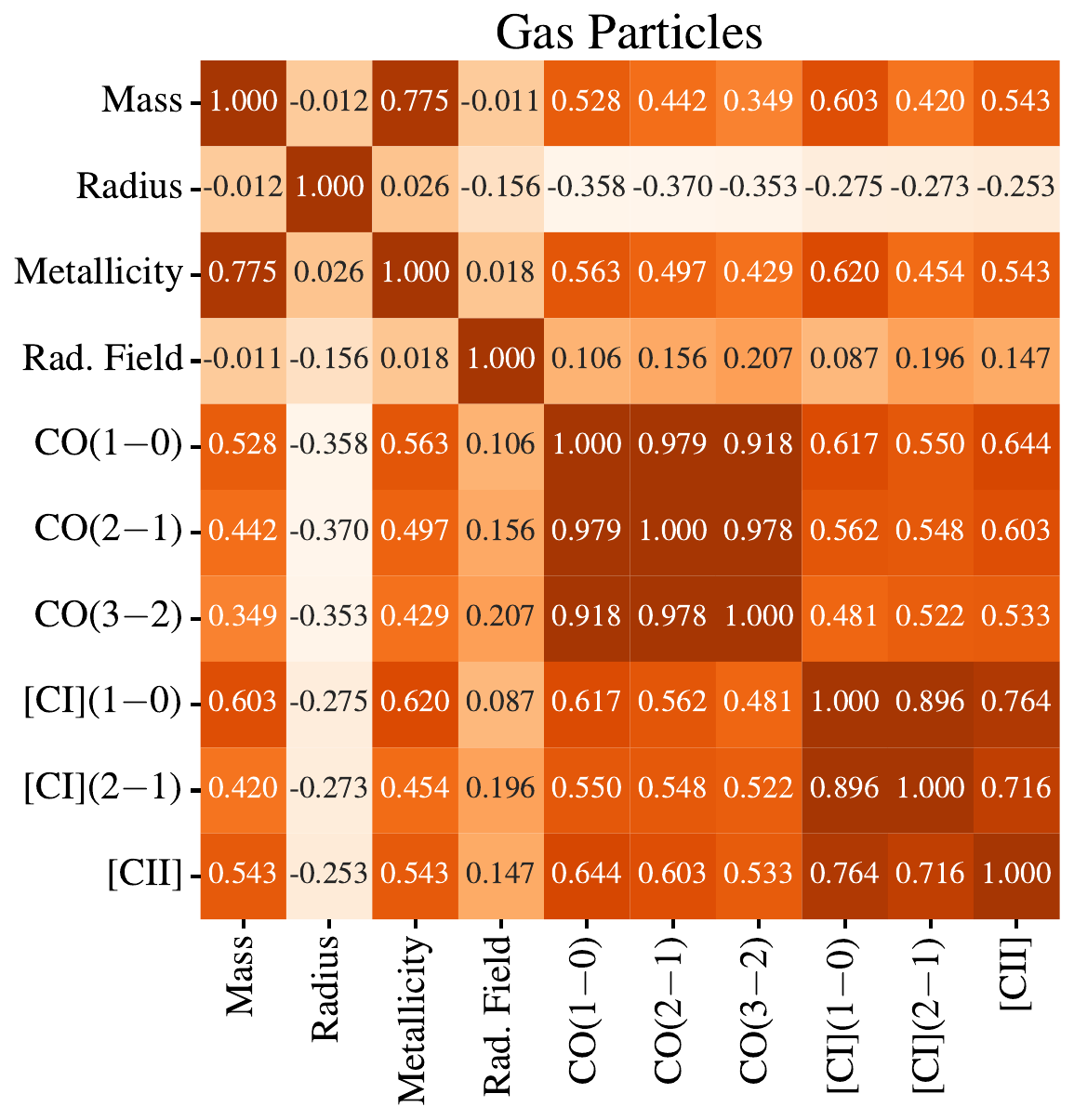}\hfil
    \includegraphics[width=0.48\linewidth]{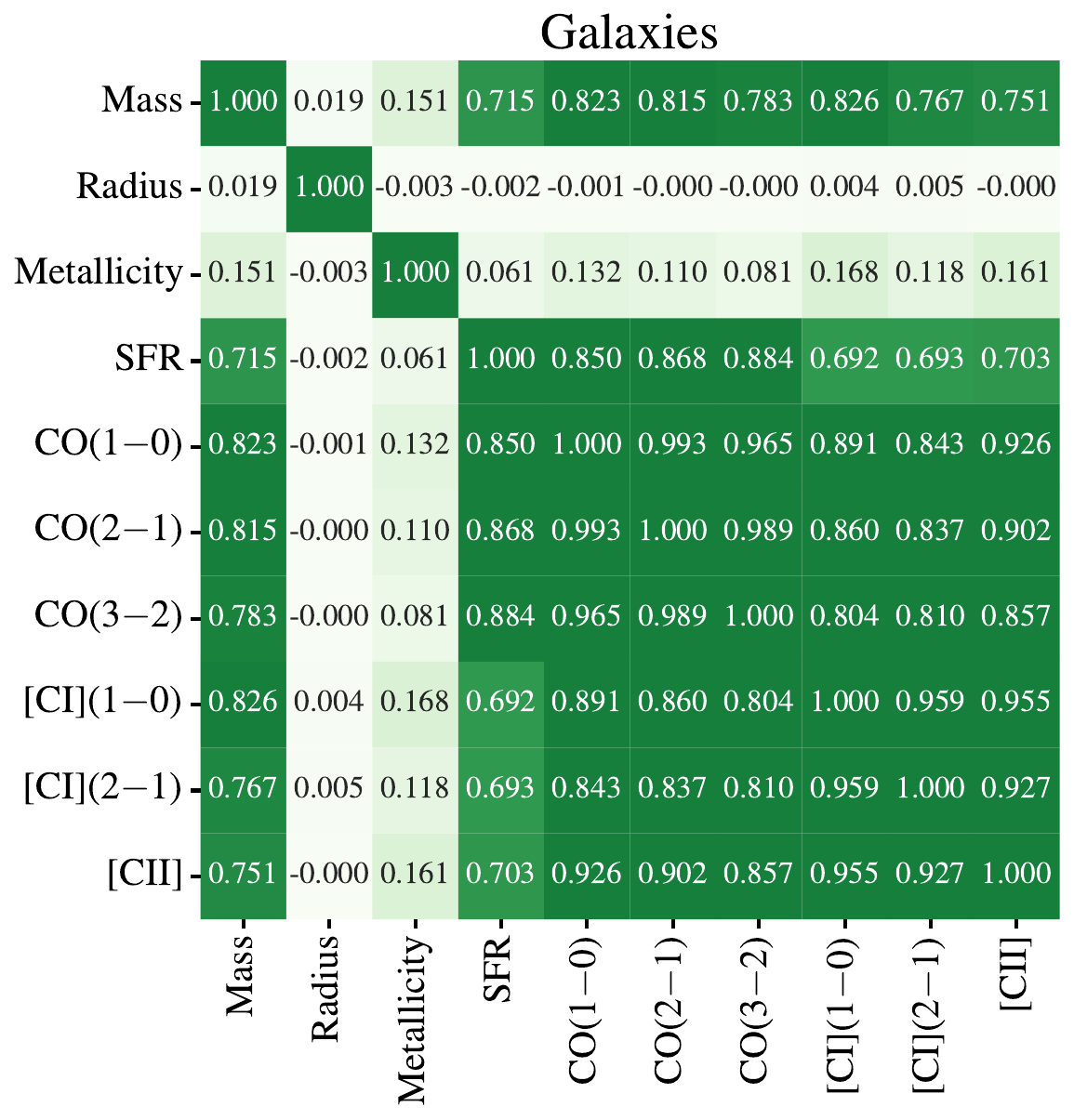}\hfil
\caption{Correlation matrices including our ML model input parameters (which vary depending on the use clouds or galaxies), and some of the lines we can predict. The values are Pearson correlation coefficients (Eq. \ref{eqn:pearson}) representing the degree of association between pairs of variables.} \textbf{Mass and metallicity are the features which affect the most all cloud line predictions. Mass and SFR are the most important when predictions are for whole galaxies.} On the left, a sample of $100,000$ gas particles. On the right, a sample of $2,500$ galaxies.
\label{fig:corr_matrices}
\end{figure*}

The model's feature importance analysis provides insights into which input parameters are most influential, as highlighted in Fig.~\ref{fig:corr_matrices}. The values within the correlation matrices represent the strength and direction of the linear relationship between pairs of variables, quantified using the Pearson correlation coefficient \citep{freedman2007statistics}. The coefficient, \( r_{xy} \), is calculated using the formula:
\begin{align}
    \label{eqn:pearson}
    r_{xy} = \frac{\rm{cov}(x,y)}{\sigma_x \cdot \sigma_y} = \frac{n(\sum xy) - (\sum x)(\sum y)}{\sqrt{[n\sum x^2 - (\sum x)^2][n\sum y^2 - (\sum y)^2]}} ,
\end{align}
where \( x \) and \( y \) are variables representing the data sets being compared, each comprising \( n \) paired data points; $\rm{cov}(x,y)$ is the covariance between $x$ and $y$; $\sigma_x$ and $\sigma_y$ are their standard deviations.

When we make predictions for gas particles, all features seem to play an important role, and line luminosities are most sensitive to mass and metallicity. In the case of predictions for whole galaxies, mass and SFR strongly affects all lines, whereas metallicity also plays an important role, especially for [\ion{C}{1}] and [\ion{C}{2}], though its effect is less pronounced than in individual gas particles. This difference arises because gas particles represent specific local conditions directly impacted by dominant physical processes that influence their emission characteristics. In contrast, galaxies encompass a broader array of physical processes and environmental factors, including variations in metallicity across different regions. Galaxy radius in our sample is weakly correlated with line luminosities and other physical parameters such as mass, metallicity, and SFR. This is due to the lack of a representative sample of galaxies in terms of their sizes in the 25 Mpc/h box.

\newpage

\begin{figure}%
    \centering
    \includegraphics[width=8.5cm]{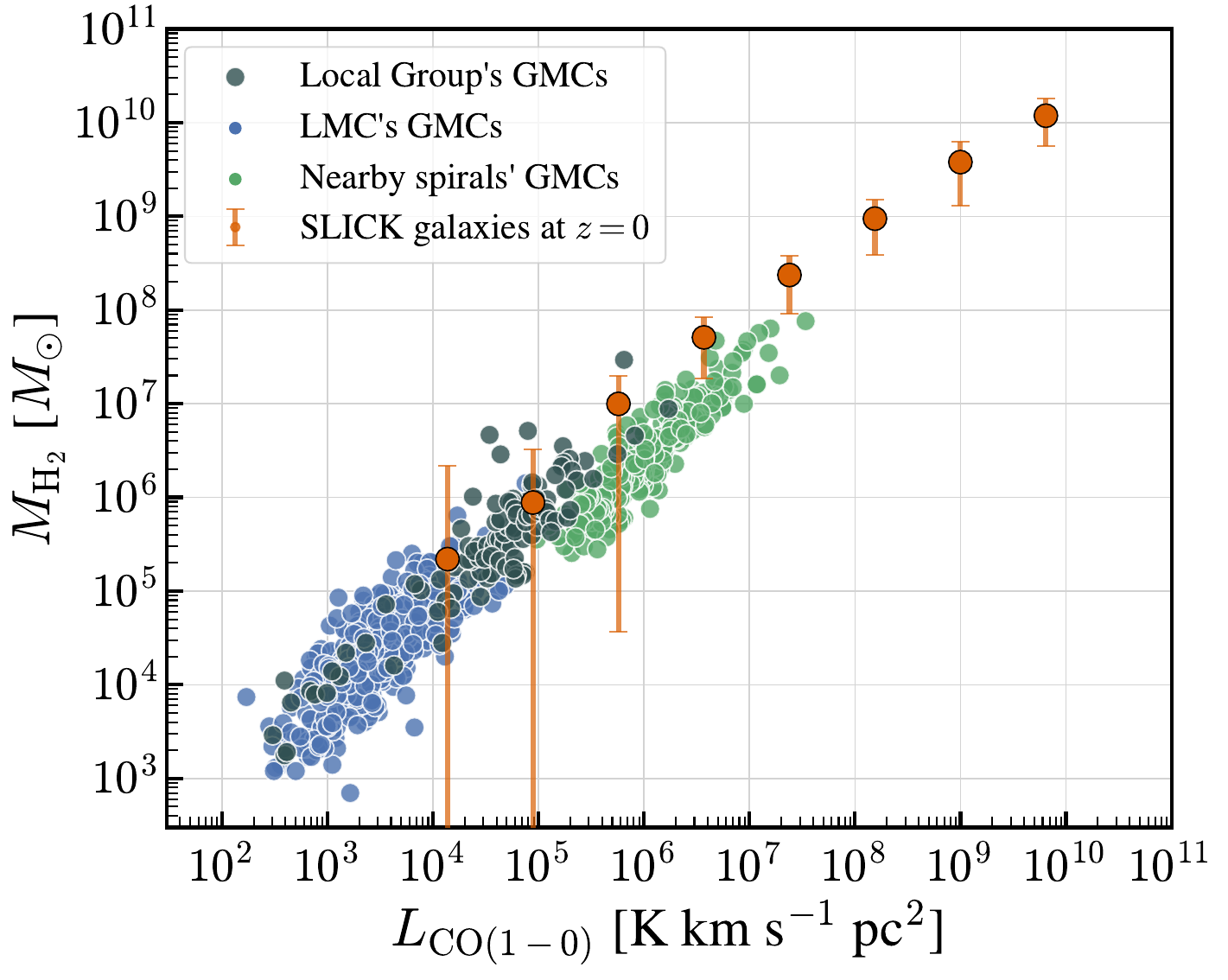}%
    \caption{\textbf{Our model ({\sc slick}) matches observational constraints for the $M_{H_2}$--$L_{[CO(1-0)]}$ relation.} In orange, {\sc slick}'s $z=0$ galaxy H$_2$ mass mean (points) and standard deviation (error bars) for different CO(1--0) luminosity bins; in other colors, observational data of extragalactic GMCs, relating their virial masses to their measured luminosities. LMC's data is from \cite{pineda_influence_2009} and \cite{wong_magellanic_2011}; nearby spirals' data is from \cite{meyer_resolved_2011}, \cite{meyer_resolved_2013}, and \cite{rebolledo_giant_2012}; and Local Group's data is from \cite{bolatto_resolved_2008}.}%
	\label{fig:CO1}%
\end{figure}

\begin{figure}%
    \centering
    \includegraphics[width=8.5cm]{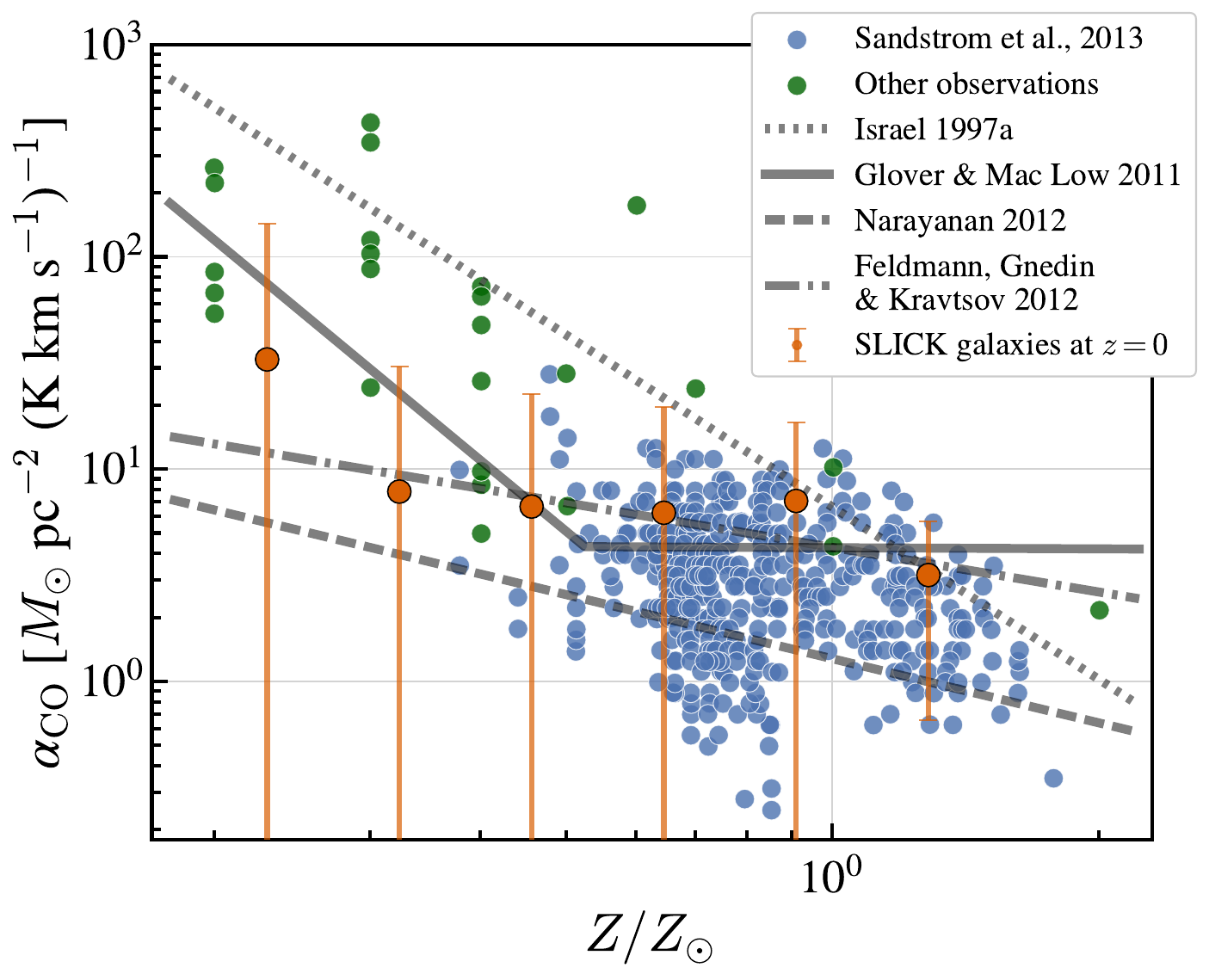}%
    \caption{\textbf{Our model ({\sc slick}) is in agreement with observational constraints for the relationship between $\alpha_{CO}$ and metallicity, and agrees with other theoretical predictions.} In orange, {\sc slick}'s z = 0 galaxy $\alpha_{CO}$ mean (points) and standard deviation (error bars) for different metallicity bins; while the other colors represent observations of nearby galaxies \citep{sandstrom_co--h2_2013,madden_c_1997,leroy_spitzer_2007,leroy_co--h2_2011,gratier_molecular_2010,roman-duval_physical_2010,bolatto_state_2011,smith_herschel_2012}. \cite{bolatto_co--h2_2013} explains the data in more detail. The solid lines represent a selection of theoretical predictions for this relation \citep{narayanan_general_2012,feldmann_x-factor_2012,israel_h2_1997,glover_understanding_2011}}.%
	\label{fig:CO2}%
\end{figure}

\begin{figure}%
    \centering
    \includegraphics[width=8.4cm]{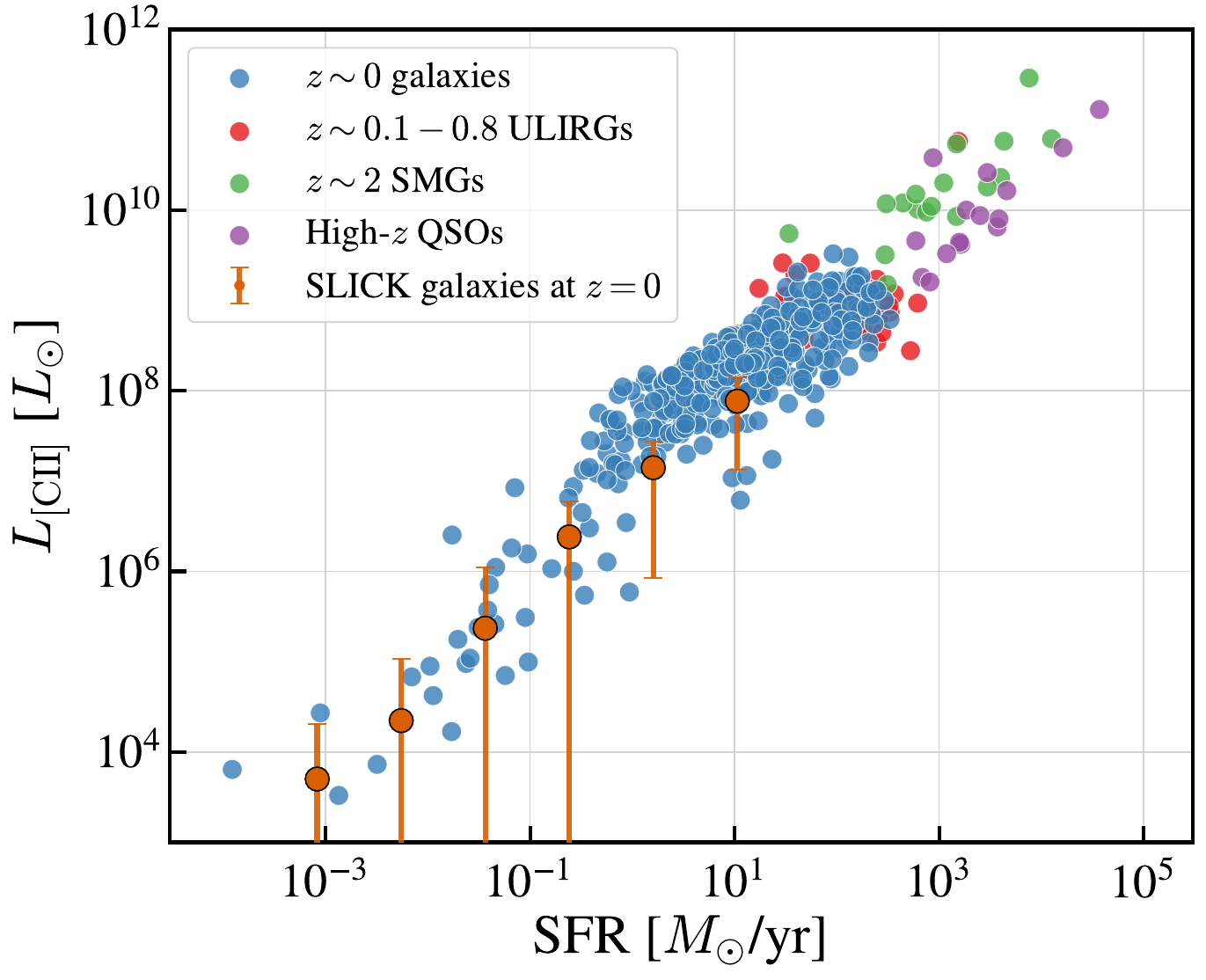}%
    \caption{\textbf{Our model ({\sc slick}) matches observational constraints for the $L_{[\mathrm{CII}]}$--SFR relation.} In orange, {\sc slick}'s galaxy [\ion{C}{2}] luminosity mean (points) and standard deviation (error bars) for different SFR bins; in other colors, observational data from \cite{brauher_compendium_2008}, \cite{diaz-santos_strikingly_2015}, \cite{farrah_far-infrared_2013}, \cite{gracia-carpio_far-infrared_2011}, \cite{rigopoulou_herschel_2014}, and \cite{swinbank_alma_2012}.}%
	\label{fig:[CII]_vs_SFR}%
\end{figure}

\subsection{Validation Tests} \label{sec:validation}

We compare our results to observations and other models in the literature, using the $z = 0$ snapshots of the {\sc Simba} simulations. Following the fluxogram in Fig.~\ref{fig:slick2}, we computed CO and [\ion{C}{2}] luminosities of half of the $z=0$ gas particles using our sub-resolution model and the radiative transfer code {\sc despotic}, and the other half using an RF model trained on 100,000 random particles from different snapshots.

First, we compare our simulated galaxy CO(1--0) luminosities and H$_2$ masses to extragalactic GMC luminosities and virial masses, as shown in Fig.~\ref{fig:CO1}. Observational data were extracted from \cite{pineda_influence_2009}, \cite{wong_magellanic_2011}, \cite{meyer_resolved_2011}, \cite{meyer_resolved_2013}, \cite{rebolledo_giant_2012}, and \cite{bolatto_resolved_2008}. Because the {\sc slick} predictions are for entire simulated galaxies, whereas observations are of individual GMCs, the mass range covered is not the same. Our least massive galaxies overlap with spiral galaxies and Local Group GMCs, generally presenting a higher molecular mass average for the same $L_{\mathrm{CO}}$ of the observed GMCs. Nevertheless, the general trend in the $M_{\mathrm{H}_2}-L_{\mathrm{CO}}$ relation is consistent.

Fig.~\ref{fig:CO2} shows the CO-to-H$_2$ conversion factor, $\alpha_{\rm{CO}}$ (defined as $\alpha_{\rm{CO}} = M_{\rm{H}_2}/L_{\rm{CO}}^\prime$) as a function of the mass-weighted metallicity of galaxies. {\sc slick}'s predictions are compared to observations \citep{sandstrom_co--h2_2013,madden_c_1997,leroy_spitzer_2007,leroy_co--h2_2011,gratier_molecular_2010,roman-duval_physical_2010,bolatto_state_2011,smith_herschel_2012,israel_h2_1997,bolatto_resolved_2008,moustakas_vizier_2010}
and theoretical predictions\citep{narayanan_general_2012,feldmann_x-factor_2012,israel_h2_1997,glover_understanding_2011}. The high metallicity range of {\sc slick}'s results agree with most observational data, especially the ones by \cite{sandstrom_co--h2_2013}. On the other hand, some of the observations at lower metallicities yield higher $\alpha_{\mathrm{CO}}$ values than our results. This said, there are significant uncertainties associated with galaxy $\alpha_{\rm{CO}}$ estimation from observations, especially at low metallicities, because they are based on limited empirical calibrations against the metallicity and dust-to-gas ratio.  

Fig.~\ref{fig:[CII]_vs_SFR} shows the relationship between [\ion{C}{2}] luminosity and SFR of the $z = 0$ snapshot galaxies. For comparison, we included the observational data measurements from \cite{brauher_compendium_2008}, \cite{diaz-santos_strikingly_2015}, \cite{farrah_far-infrared_2013}, \cite{gracia-carpio_far-infrared_2011}, \cite{rigopoulou_herschel_2014}, and \cite{swinbank_alma_2012}. Our model at $z=0$ is consistent with the observed galaxies at the same redshift. However, it does not capture the higher star formation rates and [\ion{C}{2}] luminosities due to the absence of larger structures within the 25 Mpc/h volume. Besides that, our simulated data points show a steeper trend than the one from higher-SFR galaxies. Although consistent with the limited observational dataset in the low-SFR range, this difference could be due to increased contributions to the [\ion{C}{2}] luminosity from diffuse atomic gas at low SFRs, which we do not model yet.

\subsection{Light cone}

We build our light cones based on the code methodology developed in \cite{lovell_reproducing_2021}. Initially, a predefined sky area is assumed, which sets the comoving distance covered by the light cone in each snapshot. Depending on the chosen sky area and the redshift of the snapshot, the simulation volume might be too small to cover the observed area. To address this problem, we choose a random line-of-sight alignment axis, and randomly translate the volume along the plane of the sky direction. Then we stitch each consecutive snapshot along this line-of-sight.

We show light cones for different lines spanning $z=0$ to $z=10$ in Fig.~\ref{fig:lc}. We use an area $A = 0.5$ deg$^2$, and paint CO(1-0) (top panel), [\ion{C}{1}](1-0) (middle panel), and [\ion{C}{2}] (bottom panel) luminosities on each galaxy of the light cone. To paint all 504,420 galaxies of the light cone we used our RF ML model trained on 4,000 galaxies. As we run the full-RT version of {\sc slick} on more clouds/galaxies, we will increase our training dataset and improve predictions.

\begin{figure*}[htb!]
\centering
    \includegraphics[width=0.87\linewidth]{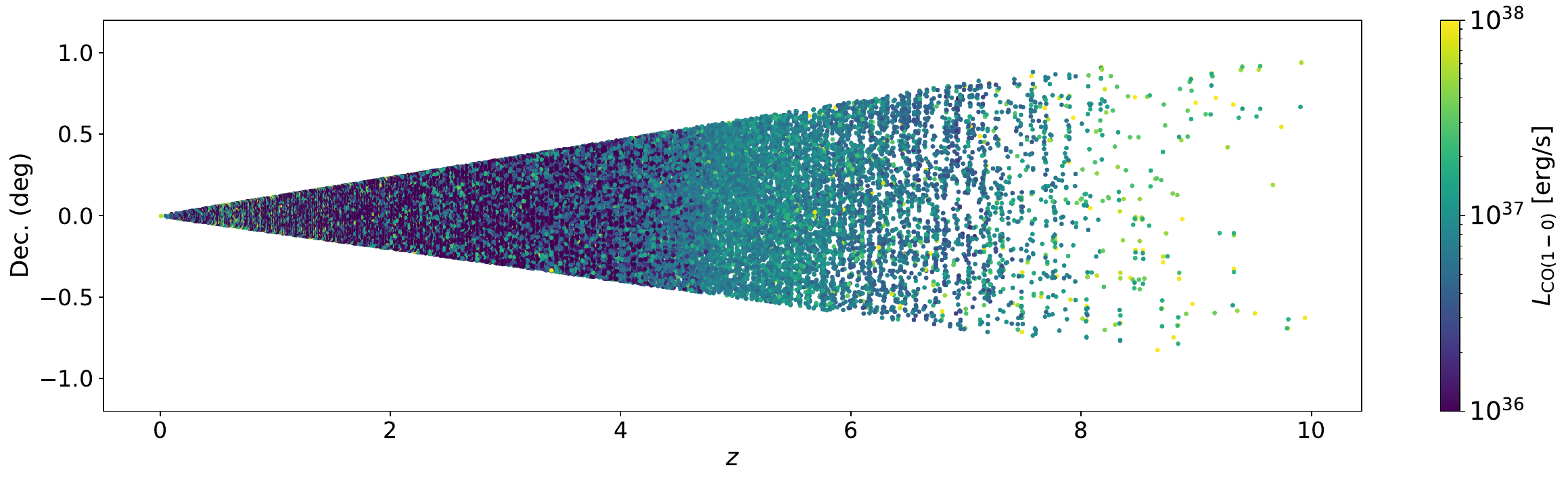}\hfil
    \includegraphics[width=0.87\linewidth]{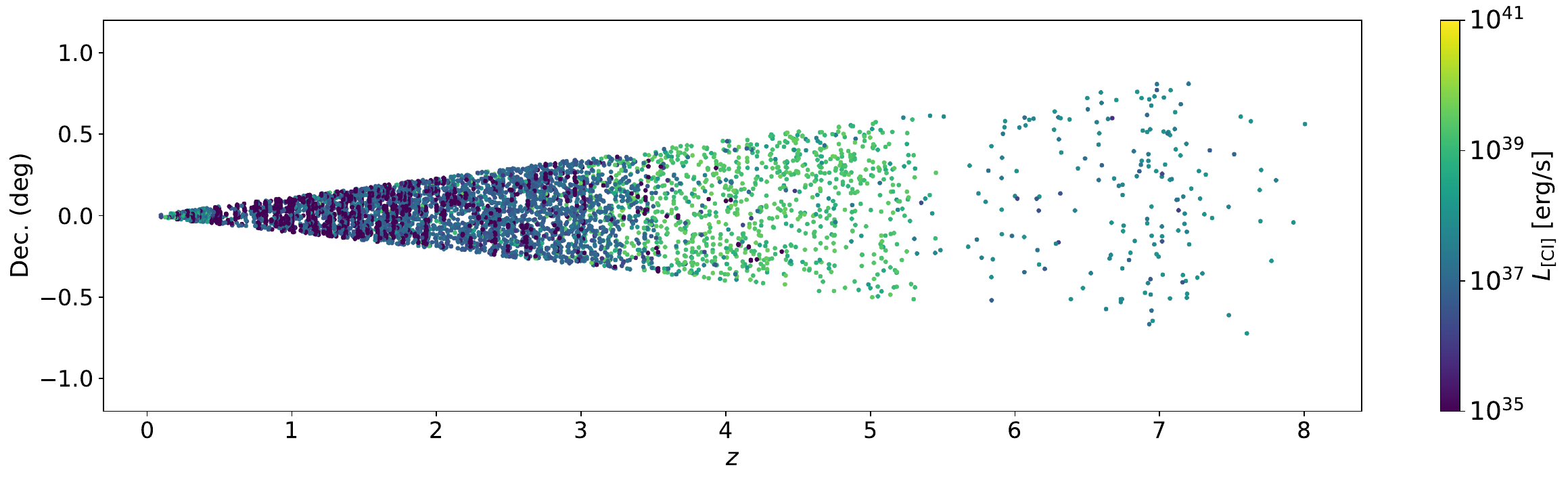}\hfil
    \includegraphics[width=0.87\linewidth]{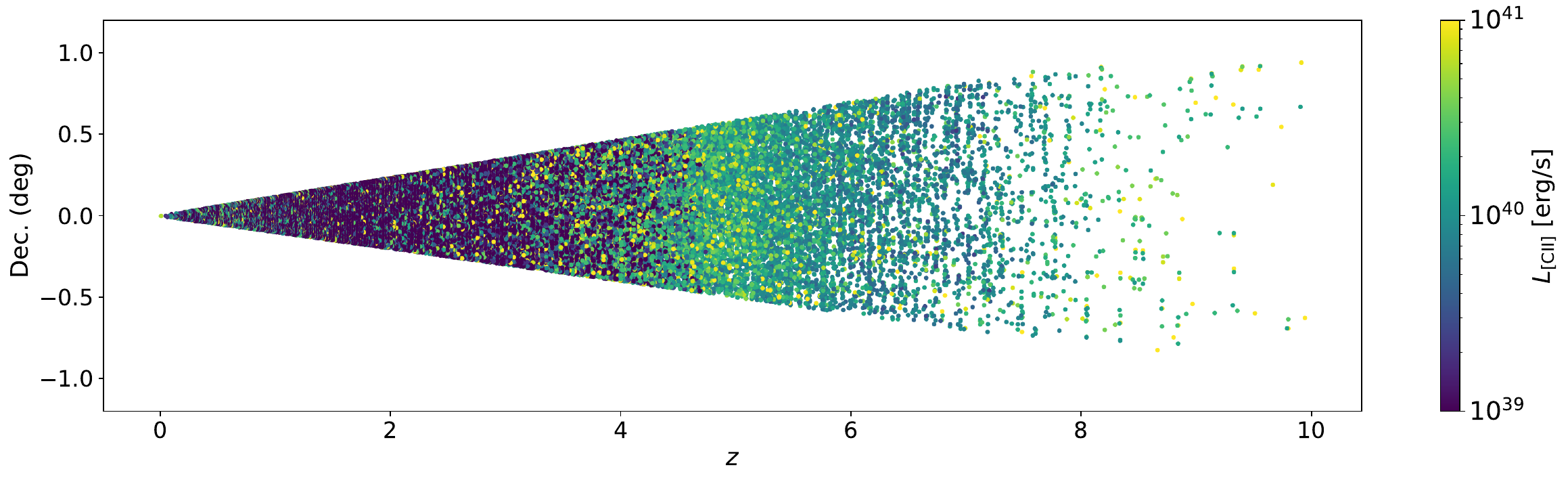}\hfil
    \caption{\textbf{Our combination of RT-based cloud luminosity calculations and ML modeling allows for the construction of entire molecular line light cones.}
    Here we show {\sc Simba} + {\sc slick} light cones for CO(1--0) (top), [\ion{C}{1}](1--0) (middle), and [\ion{C}{2}] (bottom) over 0.5 deg$^2$, between 0.1 $\leq$ $z$ $\leq$ 10. Each point represents a galaxy colored by their line luminosities. The x-axis shows the redshift range, and the y-axis shows the declination in degrees.}%
    \label{fig:lc}%
\label{fig:lc}
\end{figure*}

\subsection{Luminosity Functions}

\begin{figure}
    \centering
    \includegraphics[width=8.5cm]{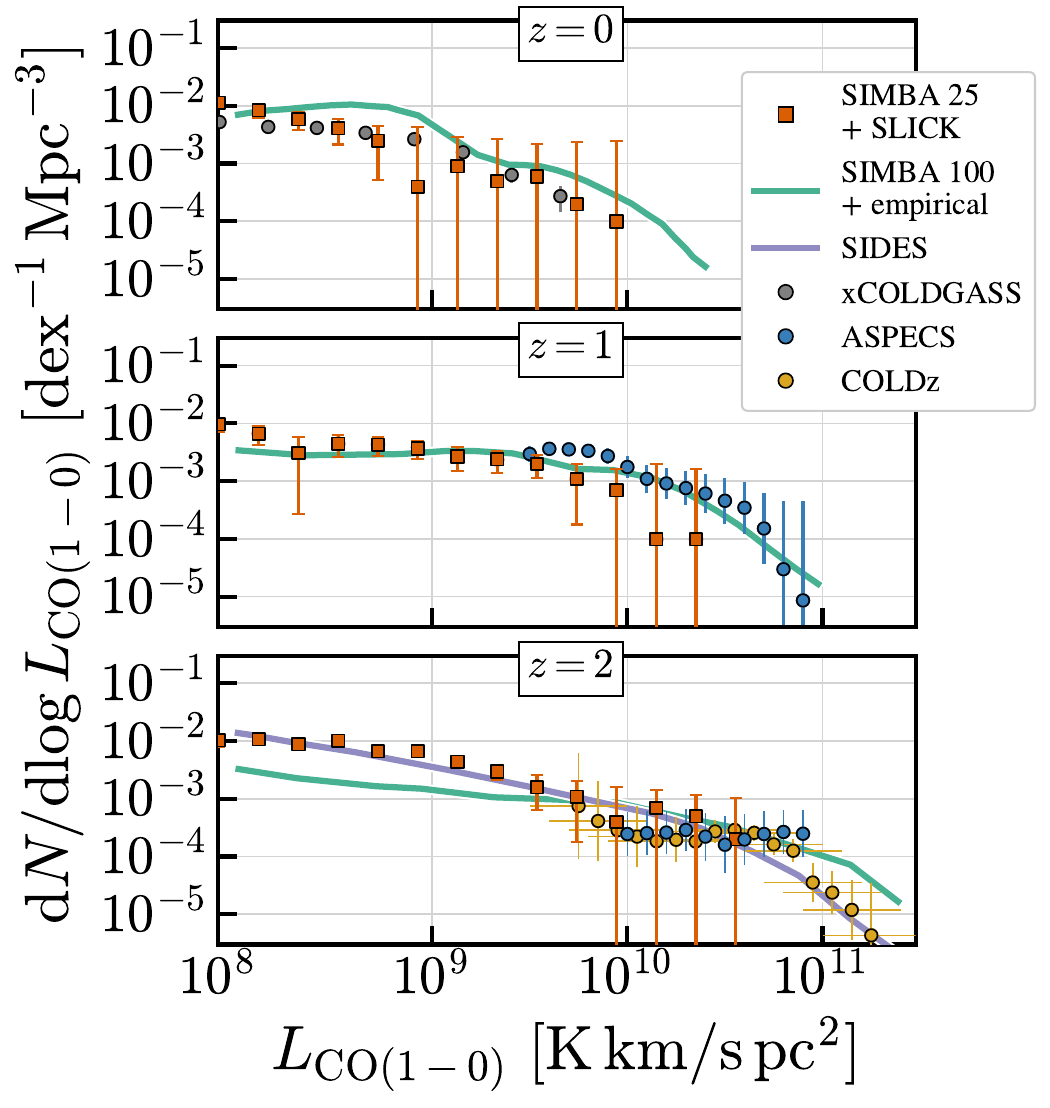}%
    \caption{\textbf{{\sc slick}'s CO(1-0) luminosity functions in agreement with observations.} CO(1–0) luminosity functions at $z = 0$, 1, and 2. In orange, results from {\sc slick} (applied on {\sc Simba}'s 25 Mpc/h simulation). The green and purple lines refer to simulations using {\sc Simba} 100 Mpc/h, with luminosities computed using the \cite{narayanan_general_2012} prescription for $\alpha_{\mathrm{CO}}$, and the {\sc SIDES} simulations, respectively. The gray, red, and blue circles are observations from xCOLD GASS \citep{saintonge_xcold_2017}, ASPECS \citep{aravena_alma_2019}, and COLDz \citep{riechers_coldz_2019}. {\sc slick} reproduces the observational data at $z=0$ and $z=2$ better than {\sc Simba} and SIDES.}%
	\label{fig:lumfunc}%
\end{figure}

We compute CO(1–0) luminosity functions using the $z=0$ dataset described in Section \ref{sec:validation}, and the full-RT form of {\sc slick} run on all gas particles of the $z = 1$, and $2$ 25 Mpc/h {\sc Simba} simulation. Fig.~\ref{fig:lumfunc} shows the CO(1-0) luminosity functions for those redshifts, where we compare our results to observations and other simulation-based models. $z\sim 0$ observational data is extracted from xCOLD GASS \citep{saintonge_xcold_2017}, $z=1$ from ASPECS \citep{aravena_alma_2019}, and $z=2$ COLDz \citep{pavesi_co_2018,riechers_coldz_2019}. We show \cite{dave_galaxy_2020}'s luminosity functions in all $z = 0,1$, and $2$, where they used the $100$ Mpc/h {\sc Simba} simulation, but estimated luminosities based on empirical relations. For $z=2$ we also include SIDES simulations by \cite{bethermin_concerto_2022}, where SFRs and luminosities also come from empirical estimations. Error bars for {\sc slick}'s data account for sample variance in each luminosity bin. {\sc slick}'s CO(1-0) luminosity functions are in agreement with the empirical models, and most of the observational data in all redshifts lie within {\sc slick}'s error bars. Applying {\sc slick} to larger boxes will allow for more precise measurements at higher luminosities.

\begin{figure}
    \centering
    \includegraphics[width=8.5cm]{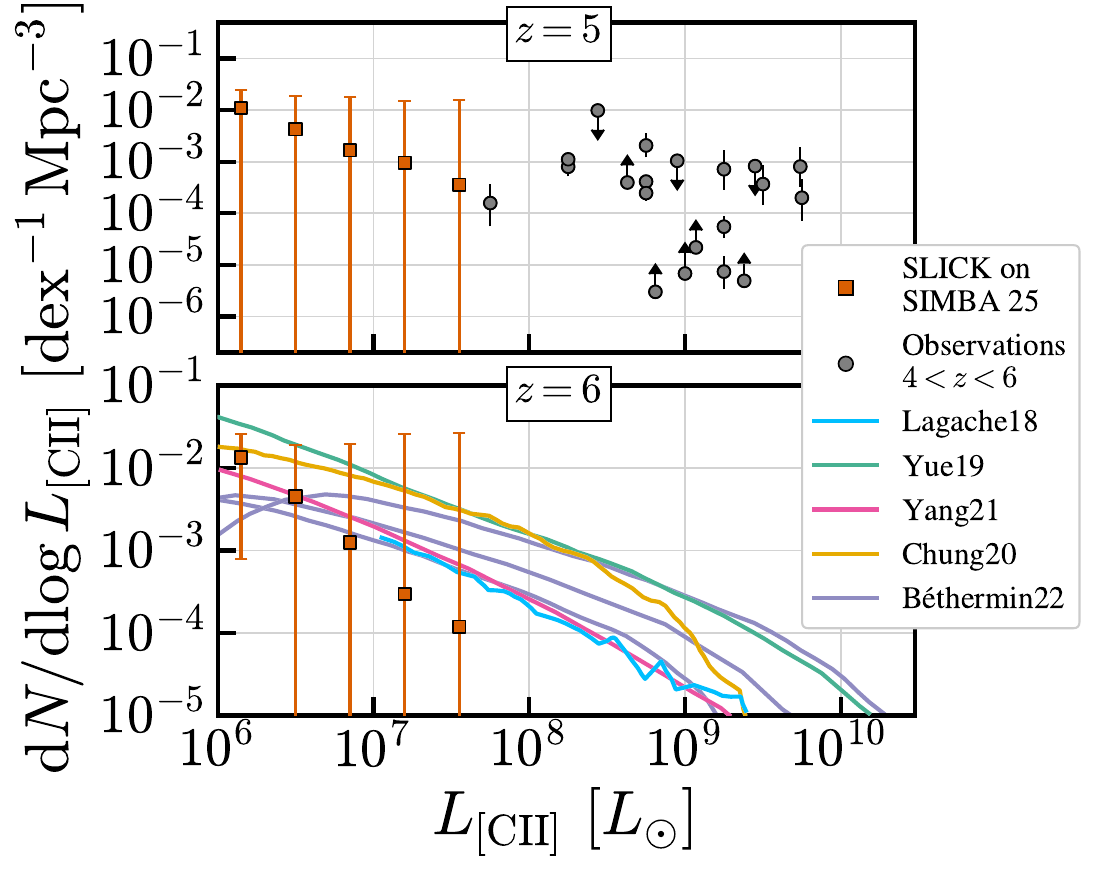}%
    \caption{\textbf{{\sc slick}'s [\ion{C}{2}] luminosity functions for high redshifts puts constraints on low luminosity end predictions.} The upper panel shows {\sc slick}'s ML results for $z=5$, and compares it to ALPINE observations \citep{yan_alpine-alma_2020, loiacono_alpinealma_2021}, and observations by \cite{swinbank_alma_2012,capak_galaxies_2015,yamaguchi_blind_2017,cooke_alma_2018}, all with $4<z<6$. Our current simulation volume does not reach the higher luminosity structures. The lower panel shows {\sc slick}'s results for $z=6$, which is in agreement within the error bars to other simulations \citep{lagache_cii_2018,yue_studying_2019,yang_multitracer_2021,chung_forecasting_2020,bethermin_concerto_2022}.}%
	\label{fig:lumfunc_cii}%
\end{figure}

Because upcoming [\ion{C}{2}] line intensity mapping observations will focus on $z > 5$, especially closer to the reionization epoch, we show in Fig.~\ref{fig:lumfunc_cii} [\ion{C}{2}] luminosity functions for redshifts $z=5$ and $z=6$. Different from the full RT calculations for the CO(1-0) luminosity functions, we predict the [\ion{C}{2}] line luminosities using our RF model trained on a set of $70,000$ gas particles picked from the $z=0,1,2,8$, and $10$ simulation snapshots. The top panel of Fig.~\ref{fig:lumfunc_cii} compares {\sc slick}'s $z=5$ results to ASPECS \citep{yan_alpine-alma_2020,loiacono_alpinealma_2021} and other $4<z<6$ observations \citep{swinbank_alma_2012,capak_galaxies_2015,yamaguchi_blind_2017,cooke_alma_2018}. The comparison here is complicated by the fact that the observational data remains sparse, and the 25 Mpc/h box does not contain the large structures required to probe the high luminosity end. However, the overall trend of our results is in agreement with observations. The bottom panel of Fig.~\ref{fig:lumfunc_cii} shows {\sc slick}'s $z=6$ results compared to other models in the literature \citep{lagache_cii_2018,yue_studying_2019,yang_multitracer_2021,chung_forecasting_2020,bethermin_concerto_2022}. At $z=6$ the box size limitation is even more prominent; so far {\sc slick}'s results agree with the other models.

\section{Discussion} \label{sec:discussion}

In this work, we have developed the first-ever method that calculates exact CO, [\ion{C}{1}], and [\ion{C}{2}] luminosities on a cloud-by-cloud basis in full cosmological hydrodynamical simulations, with flexibility to expand to larger boxes through ML. We presented an application of {\sc slick} to the {\sc Simba}'s 25 Mpc/h simulation, but we reiterate that this method can be applied to any simulation or galaxy formation model. It has already been used on IllustrisTNG datasets, for which results will come in a forthcoming paper.

From the modeling point of view, there are limitations in our method that can be explored further in future projects. First, {\sc Simba} cosmological simulations (as well as many other state-of-the-art cosmological hydrodynamical simulations) enforce some form of a temperature (or pressure) floor in order to maintain a stable multiphase ISM. In our model, this impacts the particle radius that we input to our model to calculate line luminosities. This problem is constrained by the resolution of the simulation, and there is always a trade-off between resolution and achieving cosmological volumes. Our sub-resolution and molecular line modeling of the {\sc Simba} 25 Mpc/h simulation yields results that align well with observational constraints. To expand the box size, we will adopt solutions such as the ML approach described in Section \ref{sec:slick} using full galaxies. We can train the ML model using galaxy luminosities from higher resolution boxes and apply to higher-volume ones.

Another challenge for modeling line emission in simulations such as {\sc Simba} and IllustrisTNG is to model the radiation field around each gas particle, which is currently not performed on-the-fly. Here, we  implemented a varying FUV and cosmic-ray field, depending on gas particles neighbouring each cloud. Our model could be improved through a better understanding of the behavior of CR ionization, and by directly modeling the dust temperature.

LIM experiments require fields of view that can capture enough of the large-scale structure to constrain cosmological parameters. Thus a 25 Mpc/h {\sc Simba} simulation box is not representative, especially at higher redshifts, as shown in Fig.~\ref{fig:lumfunc_cii}. Running {\sc slick} in its full radiative transfer form on larger snapshots (such as the 300 Mpc IllustrisTNG) would be unfeasible, and the poorer resolution of such larger boxes could affect the accuracy of our luminosity calculations (although this yet needs to be tested and quantified). The ML approach implemented on {\sc slick} makes it possible to calculate luminosities in these larger boxes independent of observational empirical relations. However, this process can be improved by using more representative samples of galaxies when training the ML. In a forthcoming paper, we will broaden the training set with larger sets of input simulations.

For LIM experiment forward modeling, cosmic variance will be a challenge. Combining {\sc slick} with {\sc CAMELS} \citep[Cosmology and Astrophysics with MachinE Learning Simulations;][]{villaescusa-navarro_camels_2021} may provide a solution for that. {\sc CAMELS} is a suite of N-body and hydrodynamic simulations, generated using thousands of different cosmological and astrophysical parameters. By applying the luminosity prediction model from {\sc slick} to many different {\sc CAMELS} snapshots, we plan to reduce sample variance in the measurements, and test the sensitivity of different astrophysical/cosmological parameters. {\sc slick}'s scalability and flexibility will make it possible to explore a universe of datasets to better forecast future experiments and analyze upcoming ones.

\section{Conclusion} \label{sec:conclusion}

In this paper, we introduced {\sc slick} (the Scalable Line Intensity Computation Kit), a software package that revolutionizes our ability to calculate CO, [\ion{C}{1}], and [\ion{C}{2}] luminosities for clouds and galaxies within the context of full cosmological hydrodynamical simulations. Our method operates on a cloud-by-cloud basis, and through ML it offers flexibility to extend its application to simulations of varying scales, and even to different galaxy formation models. Leveraging our ML RF model, we have achieved precise predictions of cloud and galaxy luminosities, as substantiated by Fig.~\ref{fig:ml_clouds} and Fig.~\ref{fig:ml_gals}.

We have demonstrated {\sc slick}'s capabilities by applying it to the {\sc Simba} 25 Mpc/h simulation. As shown in Fig.~\ref{fig:CO1}, {\sc slick}'s integrated galaxy parameters generally follow the L$_{\mathrm{CO}}$--H$_2$ relation derived from GMC observations, and it agrees with observed galaxy data for the $\alpha_{\rm{CO}}$--$Z$ relation (as seen in Fig.~\ref{fig:CO2}), especially at metallicities close to solar. Our model underestimates $\alpha_{\rm{CO}}$ when compared to part of the sample of lower metallicity observations, however the sample in this regime is scarce and the estimation of $\alpha_{\mathrm{CO}}$ from observations relies on uncertain assumptions. Furthermore, our results agree with the L$_{\mathrm{[CII]}}$--SFR relation derived from $z=0$ galaxy observations (Fig.~\ref{fig:[CII]_vs_SFR}). In the regime we probe (SFR $\leq 10$) we find a steeper curve for the L$_{\mathrm{[CII]}}$--SFR relation compared to observations of high-z high-SFR galaxies. This is consistent with the available low-SFR data. However, this difference could also be due to increasing [\ion{C}{2}] emission contribution from diffuse gas in lower-SFRs, which is not modeled in this work.

Our modeled $z=0,1$ and $2$ CO(1–0) luminosity functions agree with observational data and other simulation-based models, with results depicted in Fig.~\ref{fig:lumfunc}. In the case of [\ion{C}{2}], we computed luminosity functions for $z = 5$ and $6$ with the intent of comparing it not only to observations but also to LIM predictions from the literature. Fig.~\ref{fig:lumfunc_cii} shows that the box size did not allow us to capture the high luminosity end from higher redshifts, but highlights the potential of our models if applied to larger simulations.

With {\sc slick}, we now have a quick, simple, and versatile way of combining the ISM details provided by high resolution simulations with the large scale properties from large simulation boxes, allowing for the construction of entire molecular line light cones with cosmological hydrodynamical simulations.

\begin{acknowledgements}
We would like to thank Romeel Davé for providing the luminosity functions from his paper, and Christopher Lovell for his {\sc Simba} light cone code. We are grateful to Christos Karoumpis, Matus Rybak, Azadeh Moradinezhad, Emilio Romano-Diaz, and Jiamin Hou for fruitful discussions; and to Dariannette Valentín Martínez for her insightful feedbacks on {\sc slick}. KG thanks Sidney Lower and Desmond Jeff for their help with understanding key concepts for this work; Naman Shukla for his assistance with {\sc slick}'s GitHub page; and Christopher Lam for the daily motivational stand-up meetings. KG also thanks all who helped pick (and then change) {\sc slick}'s acronym (Shivani Shah, Sidney Lower, William Schap, Quadry Chance, Dariannette Valentín Martínez, and Ben Capistrant). This work was supported by grant NSF AST-1909153.
\end{acknowledgements}



\clearpage

\bibliographystyle{mnras}
\bibliography{all}


\label{lastpage}
\end{document}